\pdfoutput=1
\documentclass[
 pra, reprint,
 superscriptaddress,
 amsmath,amssymb,
 aps,longbibliography
]{revtex4-2}

\usepackage{graphicx}
\usepackage{amsmath}
\usepackage{xcolor}
\usepackage[bookmarks=false]{hyperref}
\usepackage[normalem]{ulem}
\usepackage{soul}
\usepackage{fancyhdr}
\newcommand{\cc}[1]{{\color[rgb]{0,0,0}#1}}

\newcommand{\ket}[1]{\mathinner{|#1\rangle}}
\begin{document}

\title{Modular tunable coupler for superconducting circuits}

\author{Daniel L. Campbell}
\email{Daniel.Campbell.22@us.af.mil}
\affiliation{Air Force Research Laboratory, Information Directorate, Rome NY 13441 USA}
\author{Archana Kamal}
\affiliation{Department of Physics and Applied Physics, University of Massachusetts, Lowell, MA 01854 USA}
\author{Leonardo Ranzani}
\affiliation{Raytheon BBN Technologies, Cambridge MA 02138, USA}
\author{Michael Senatore}
\affiliation{Air Force Research Laboratory, Information Directorate, Rome NY 13441 USA}
\affiliation{Department of Physics, Syracuse University, Syracuse NY 13244-1130 USA}
\author{Matthew LaHaye}
\email{Matthew.LaHaye@us.af.mil}
\affiliation{Air Force Research Laboratory, Information Directorate, Rome NY 13441 USA}
\date{\today}
\begin{abstract}
The development of modular and versatile quantum interconnect hardware is a key next step in the scaling of quantum information platforms to larger size and greater functionality. For superconducting quantum systems, fast and well-controlled tunable circuit couplers will be paramount for achieving high fidelity and resource efficient connectivity, whether for performing two-qubit gate operations, encoding or decoding a quantum data bus, or interfacing across modalities. Here we propose a versatile and internally-tunable double-transmon coupler (DTC) architecture that implements tunable coupling via flux-controlled interference in a three-junction dcSQUID. Crucially, the DTC possesses an internally defined zero-coupling state that is independent of the coupled data qubits or circuit resonators. This makes it particular attractive as a modular and versatile design element for realizing fast and robust linear coupling in several applications such as high-fidelity two-qubit gate operations, qubit readout, and quantum bus interfacing.
\end{abstract}
\maketitle
\section{Introduction}

Recently, several demonstrations of noisy intermediate scale quantum \emph{NISQ} processors involving complex systems comprising tens of coupled qubits, have cemented superconducting circuits as a leading platform for large scale quantum processing~\citep{cross2018ibm, arute2019quantum, gong2021quantum, ai2021exponential, wu2021strong}. Further scaling of quantum processors and the development of distributed quantum architectures will likely benefit from reconfigurable qubit connectivity, tunable dispersive readout, bus interfacing, and transducer interconnects. For all these functionalities the use of a standalone circuit interconnect, a coupler, helps to preserve the coherence of the interconnected circuits while enabling rapid, high fidelity entangling operations between them. 

Tunable couplers use an external control parameter to turn on and off an effective coupling $g_{\textrm{eff}}$. In superconducting circuits threading flux through a superconducting quantum interference device (SQuID) and applying microwave driving fields are examples of external control parameters. We highlight two broad approaches to tunable coupling: (i) couplers that use current-divider circuit elements~\citep{Niskanen2006, Niskanen2007, Allman2014, Whittaker2014, Bialczak2011, Chen2014, Geller2015, Roushan2016, Neill2018, Noh2021} (a recent well-known example is the \emph{gmon} coupler~\citep{Chen2014, Geller2015}), and (ii) couplers that interfere direct and virtual interaction pathways~\citep{Yan2018, Sung2021, Li2020, Collodo2020, Xu2020, stehlik2021, sete2021floating, marxer2022long, kandala2021demonstration, moskalenko2022high, Xu2020, Zhao2022}, such as the MIT-style single transmon coupler~\citep{Yan2018, Sung2021, Li2020, Collodo2020, Xu2020, stehlik2021, sete2021floating}. For the latter case, mutual inductance or capacitance between two qubits constitutes a direct interaction pathway while interactions mediated by coupler circuitry constitute a virtual interaction pathway. Between these two coupling approaches, current divider couplers respond more linearly with flux, lending themselves to parametric coupling applications~\citep{Zakka-Bajjani2011, Noh2021, Roushan2016}. On the other hand, inductively connecting the qubits introduces extra flux degrees of freedom which, by extension, lead to additional noise and decoherence channels, and crosstalk between local flux bias lines. A survey of coupler modalities cited in this paper, therefore, suggests that capacitive interactions lead to cleaner and potentially easier-to-control circuitry. Moreover, for circuit-QED architectures involving transmon qubits, capacitive interactions are compatible with the fixed-frequency designs routinely employed in high-coherence circuits. Therefore, on balance, capacitive rather than inductive coupling seems to provide a significant advantage for the interference- over current divider-style coupler modality especially for multi-qubit superconducting systems.

It is worthwhile to note that in the highlighted coupler approaches, the frequencies of the qubits and the magnitude of their interaction with the coupler determine the decoupling external flux bias $\Phi_e^{(0)}$ that zeroes the qubit-qubit effective coupling $g_{\textrm{eff}} = 0$; this necessitates circuit remodeling and optimization each time data qubits or architecture is even slightly modified. In addition, the quantum level structure of an interference coupler makes the realization of parametrically-driven exchange interactions challenging, an increasingly critical functionality explored in several demonstrations~\citep{Besse2020, McKay2016, Naik2017}. Motivated by these considerations several workarounds for parametric coupling, which do not involve modulating the transition frequency of the coupler itself, are gaining traction, such as modulation of the data qubit frequency itself in the presence of a fixed coupling~\citep{Strand2013, Sete2021, Reagor2018, Caldwell2018} and techniques that do not use flux driving~\citep{Niskanen2007, Majer2007, Chow2013}.

In this paper, we describe and analyze a novel double transmon coupler (DTC) design, also explored in Refs.~\cite{Goto2022, Kubo2022}, that provides the linearity to implement two-qubit parametric gates efficiently and possesses an internally defined zero-coupling state that is independent of the coupled data qubits or circuit resonators. Furthermore we introduce novel computationally friendly numerics and analytical expressions for the effective coupling of this circuit, as well as provide theoretical predictions for parametric and waveguide coupling use cases that are unique to this work.

\begin{figure*}
\begin{centering}
\includegraphics[width=\textwidth]{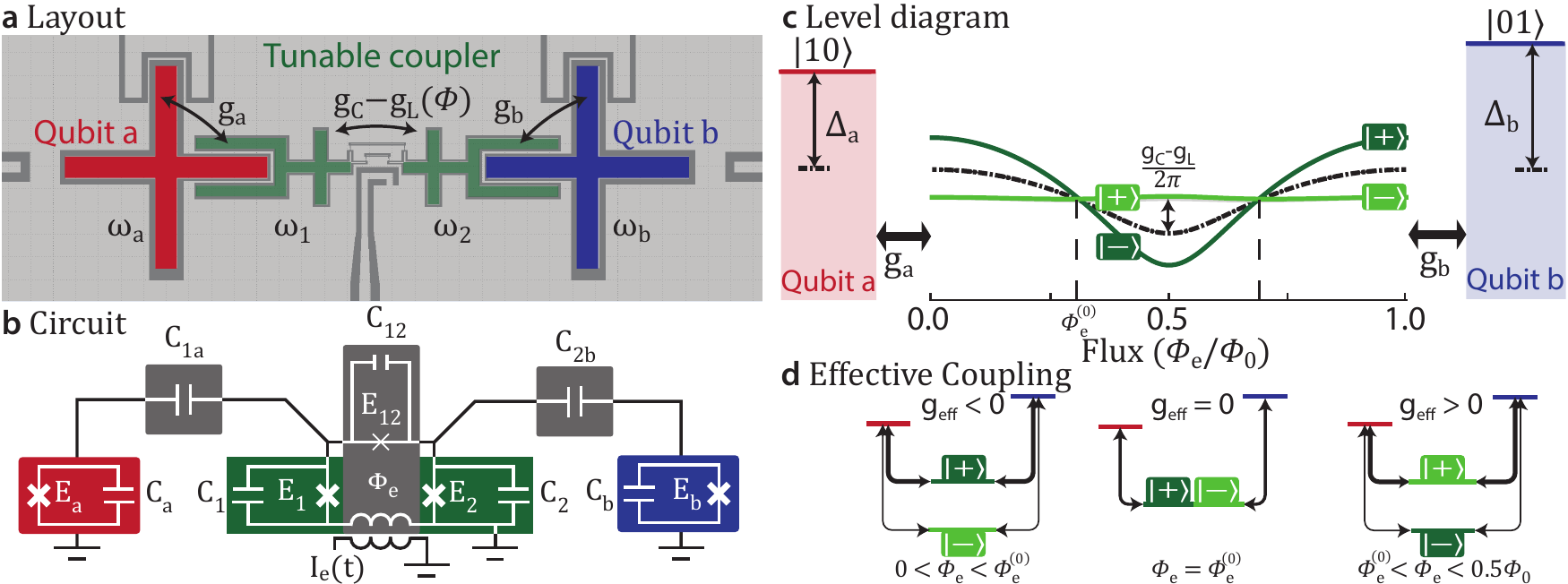}
\caption{\label{fig:couplerdispersion}
\textbf{a)} Prospective device layout. The coupler consists of a pair of transmon circuits (green) coupled by variably strong capacitance $g_C$ and strong flux tunable inductance $g_L(\Phi)$. Each coupler transmon capacitively couples, in turn, to separate qubits (red and blue). $\{\omega_a,\omega_{1},\omega_{2},\omega_b\}/2\pi$ label transmon plasma frequencies in order of appearance, left to right. 
\textbf{b)} The equivalent circuit representation used for the analysis in this work [appendix~\ref{app:circ}].
\textbf{c)} A flux-controlled interaction $g_C-g_L$ turns on a hybridization between the coupler transmons. Solid lines represent hybridized coupler eigenenergies $\omega_{\pm}$: the dark and light green lines show $\omega_{\pm}$ for capacitive coupling $g_C/2\pi = 0$~MHz and $123$~MHz, respectively. Average coupler energy, $\bar{\omega}=(\omega_1+\omega_2)/2$, is represented with the dashed black line. The detuning of each qubit from the average energy is denoted with the detuning $\Delta_j = \omega_j-\bar{\omega}$ for $j \in \{a,b\}$.
\textbf{d)} Different coupling configurations realized for three distinct flux biases. For simplicity of presentation, we have assumed $\omega_1=\omega_2$ (i.e. $\delta=0$) in (c) and (d), for which the magnitude of the interaction between coupler transmons can be directly read off from the splitting between the $|+\rangle$ and $|-\rangle$ eigenstates. 
}
\end{centering}
\end{figure*} 

The DTC uses two inductively coupled transmon qubits to combine the linearity of current divider couplers with the capacitive coupler-qubit interactions of interference-style couplers at the cost of added level structure. From this combination of attributes, we predict fast, linear, and low noise tunable coupling for the DTC that is compatible with both fixed-frequency transmon architectures and parametric coupling use cases. The design paradigm works on a simple principle: each qubit capacitively interacts with a different transmon belonging to the coupler, as illustrated in Figs.~\ref{fig:couplerdispersion}a-b. These two coupler transmons in turn hybridize with an interaction that can be flux-tuned from negative to positive values. The qubits thus virtually interact via simultaneous coupling to the same hybridized states. Moreover, this net virtual interaction between the qubits can be turned off by flux-tuning the coupler transmon interaction and hybridization to zero. 

Crucially, the addition of a second interposing transmon in the DTC compared to the interference approach suppresses the effective capacitance coupling between data qubits. The suppression geometrically scales with each interposing coupler transmon in the dispersive limit. By controllably hybridizing the DTC's transmons, effective coupling comparable to that of a single interposing transmon is achievable, while preserving the increased data qubit isolation. Moreover, the junction and capacitor parameters of the DTC set the functional dependence of the DTC's flux-tuning almost exclusively, making the decoupling flux bias $\Phi_e^{(0)}$ depend only weakly on the transition frequencies of the qubits or qubit-DTC interaction strength (if at all). In summary, this \textit{hybridization style} coupler isolates qubits in its `off' configuration and its internal operations are insensitive to the frequencies and use case of the qubits to which it's coupled: a desirable feature of modular design. We will also see that the additional transmon degree of freedom in the DTC can be leveraged to realize parametric coupling that mitigates deleterious nonlinear effects such as rectification in the presence of fast flux modulation.

\section{Double Transmon Coupler Model and Operation}

The DTC circuit consists of two superconducting transmon qubits~\cite{Koch2007} sharing a common ground. The Josephson and charging energies for each transmon are denoted by $E_{j}$ and  $E_{Cj}$ for $j\in \{1,2\}$. A relatively high inductance junction $E_{12} \sim 0.1 \times E_{1,2}$ connects the two transmons, forming a three junction dc SQuID, see Fig.~\ref{fig:couplerdispersion}b. A local bias line with an externally sourced current threads a tunable magnetic field through the SQuID, which tunes the inductive interaction between the transmons,
\begin{equation}
g_L = \frac{4E_{12}'\sqrt{E_{C1}E_{C2}}}{\hbar^2\sqrt{\omega_1\omega_2}}, 
\end{equation}
where  $E_{12}'\sim E_{12}\cos{(\Phi_e/\varphi_0)}$ is the external flux-dependent Josephson energy of the center junction [appendix~\ref{app:circ}] and $\varphi_0 = \Phi_0/2\pi$ is the reduced flux quantum. The plasma frequencies, $\omega_j = (8(E_j'+E_{12}')E_{Cj})^{1/2}$ also tune modestly with flux through flux-dependence of both $E_{12}'$ and $E_j'\sim E_j-E_{12}^2\sin^2{(\Phi_e/\varphi_0)}/2E_j$. In contrast, the capacitive interaction between the coupler transmons,
\begin{equation}
    g_C \approx \frac{C_{12}\sqrt{\omega_1\omega_2}}{2\sqrt{(C_1+C_{12})(C_2+C_{12})}},
\end{equation}
remains relatively constant as a function of flux. These inductive and capacitive contributions compete to define the net exchange interaction $g_C - g_L(\Phi_e)$, as shown in Fig.~\ref{fig:couplerdispersion}c [also see Eq.~(\ref{eq:HOfull})].
\par
To model and simulate this coupling approach, we now consider a chain of four transmon circuits. The outer transmons operate as data qubits with plasma frequencies $\{\omega_{a}, \omega_{b}\}$, while the inner transmons comprise the coupler with plasma frequencies $\{\omega_{1}, \omega_{2}\}$. Each qubit capacitively couples to a \textit{different} coupler transmon, with strength $g_{a}$ and $g_{b}$ respectively. The effective coupling between the data qubits may then be understood as a competition between virtual interactions mediated through the $|+\rangle$ and $|-\rangle$ coupler eigenstates, as shown in Fig.~\ref{fig:couplerdispersion}(d). This allows tuning $g_{\textrm{eff}}$ from positive to negative values as a nearly linear function of $g_C-g_L(\Phi_e)$, and guarantees the existence of an external flux $\Phi_e^{(0)}$ such that $g_{\textrm{eff}} = 0$. Anharmonicity in the DTC renormalizes $g_{\textrm{eff}}$ at the $20\%$ level but otherwise the form for $g_{\textrm{eff}}$ and $\Phi_e^{(0)}$ is similar to that of coupled harmonic oscillators with the same connectivity. The equivalent harmonic oscillator Hamiltonian of circuit can then be represented as [see appendix~\ref{app:full}],
\begin{align}
    H/\hbar &= \sum_{j = a,b,1,2}\omega_j a^{\dagger}_ja_j
    + (g_C-g_L(\Phi_e)) (a_{1}^{\dagger}a_{2}+a_{1}a_{2}^{\dagger})\nonumber\\
    & \quad + g_a (a_{a}^{\dagger}a_{1}+a_{a}a^{\dagger}_{1})+ g_b (a_{b}^{\dagger}a_{2}+a_{b}a^{\dagger}_{2}),
    \label{eq:H1}
\end{align}
where $a^{\dagger}_j$ and $a_j$ are the raising and lowering operators, while the index $j$ identifies the respective qubit $j=\{a,b\}$ or DTC $j=\{1,2\}$ transmon modes. 
\par
To compute $g_{\textrm{eff}}$ we first diagonalize the coupler part of the circuit, yielding $\omega_{\pm} = \bar{\omega} \pm (\delta^2 + (g_C-g_L)^2)^{1/2}$ where $2\bar{\omega} = \omega_1 + \omega_2$ and $2\delta = \omega_1-\omega_2$ are the coupler transmon sum and difference plasma frequencies, respectively. Since $\{\omega_1,\,\omega_2\}$ move together with flux, the common-mode frequency $\bar{\omega}$ also tunes with flux, whereas $\delta$ may tune only weakly, if at all. With $\delta=0$, the modulation of both $g_L$ and $\bar{\omega}$ are nearly equal so that one of the two coupler eigenmode plasma frequencies tunes only weakly with flux, as shown in Fig.~\ref{fig:couplerdispersion}c. In the other regime, when $\delta\gg |g_C-g_L|$, both coupler eigenmode plasma frequencies modulate similarly with flux, driven by flux tuning of $\bar{\omega}$.
\par
\subsection{Derivation of the effective coupling}
\begin{figure}
\begin{centering}
\includegraphics[width=\columnwidth]{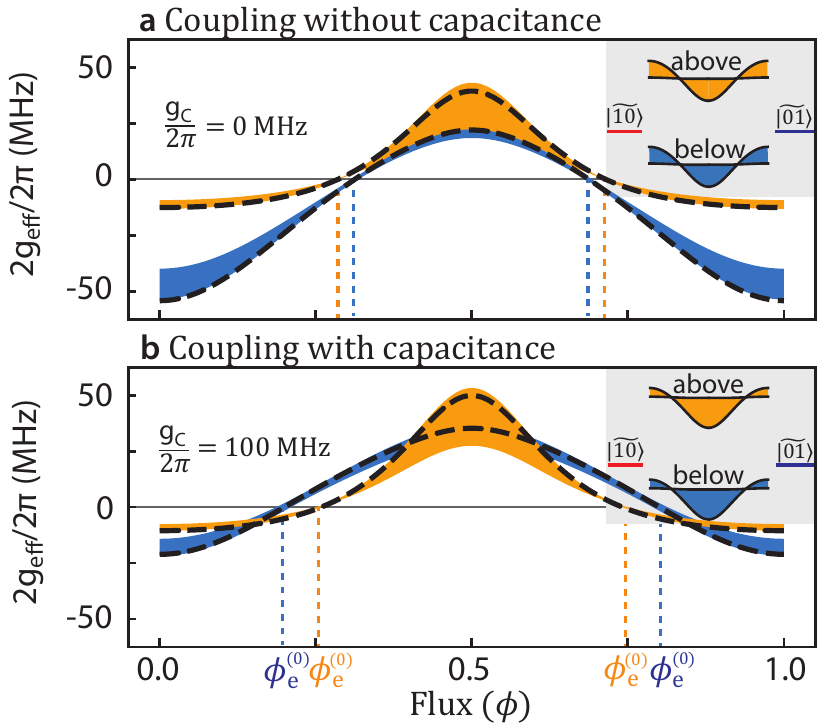}
\caption{\label{fig:effectivecoupling} The $2g_{\textrm{eff}}$ between degenerate qubits, mediated by the coupler as a function of flux, changes if the coupler transitions are set above verses below the qubit transitions, all else being equal. The shading indicates the standard deviation from the mean after applying 6\% relative Gaussian noise on all the coupler junctions, obtained using numerical simulation [Eq.~(\ref{eq:HOfullfull})]. \cc{The numerical simulations find $2g_{\textrm{eff}}/2\pi$ by taking the minimum difference between the eigenfrequencies that best correspond to the states $\ket{1000}$ and $\ket{0001}$ as $\omega_b$ is swept past $\omega_a$}. A comparison of panel (\textbf{a}) versus panel (\textbf{b}) shows that adding capacitance shifts $g_{\textrm{eff}}$ to more positive values. For each scenario, the approximate analytic $g_{\textrm{eff}}$ is plotted in black [Eq.~(\ref{app:effg w anh})]. At $\Phi_e^{(0)}$, the coupling parameters are given by $g/2\pi = 0.25\text{ GHz}$, $|\Delta|/2\pi = 1.1\text{ GHz}$, $\delta \approx 0$, and the qubit plasma frequencies are $\omega/2\pi = 4.7\text{ GHz}$.  The average junction parameters for (\textbf{a}) are $E_{12}/h = 1.46\text{ GHz}$ and $E_{1}/h,\, E_{2}/h= 7.5\text{ GHz}$, whereas for (\textbf{b}) they are $E_{12}/h = 2.56\text{ GHz}$ and $E_{1}/h,\, E_{2}/h= 23\text{ GHz}$. For all calculations, we chose $E_{12}/h$ to set the maximum inductive coupling to $g_L/2\pi = 0.31\text{ GHz}$.}
\end{centering}
\end{figure}

We now consider the data qubit interaction with the coupler eigenmodes. To gain insight into the nature of the coupler operation, we will treat this under the dispersive approximation, i.e. $g_j/|\omega_j-\omega_{\pm}| < 0.1$, where approximate analytical expression can be derived. To this end, the first step involves perturbatively decoupling the coupler eigenmodes from the qubit modes using Schrieffer-Wolff transformation of the form  
$U_{SW} = \text{exp}{\left[\sum_{\pm}\sum_{j=a,b}\frac{g_j}{\omega_j-\omega_{\pm}}(a^{\dagger}_ja_{\pm}-a_ja^{\dagger}_{\pm})\right]}$. Retaining terms to second order in $g_j/(\omega_j-\omega_{\pm})$, and continuing to omit qubit and DTC anharmonicity, leads to the following effective two-qubit Hamiltonian, 
\begin{align}
    \tilde{H}/\hbar \approxeq \sum_{j=a,b}(\omega_j+\omega_{\textrm{ac}}^j)a_j^{\dagger}a_j + g_{\textrm{eff}}\left(a^{\dagger}_a a_b+a_aa^{\dagger}_b\right),\label{eq:simpleH}
\end{align}
%
%
with a simplified form of the effective qubit-qubit coupling and dispersive shifts on the qubits given by [see appendix~\ref{app:SWT}],
\begin{align}
    &g_{\textrm{eff}} \approx \sum_{j=a,b} \frac{g_a g_b [(g_C-g_L)+(g_C+g_L)(\Delta_j/\bar{\omega})]}{2D_j^2},\label{eq:corot}\\
    &\omega_{\textrm{ac}}^a \approx g_a^2\Big[\frac{\omega_a-\omega_1}{2D_j^2}\Big],\quad
    \omega_{\textrm{ac}}^b \approx g_b^2\Big[\frac{\omega_b-\omega_2}{2D_j^2}\Big],\label{eq:acStark}
\end{align}
where $D_j^2 = \Delta_j^2 - \delta^2 - (g_C-g_L)^2$. The flux dependence of $\bar{\omega}$, and therefore ${\Delta_j = \omega_j - \bar{\omega}}$, accounts for the non-sinusoidal tuning with flux shown in Fig.~\ref{fig:effectivecoupling}. Operating in the regimes where ${|\omega_a-\omega_1|,\,|\omega_b-\omega_2|\gg |g_C-g_L|}$ and to a lesser extent when $\delta\gg |g_C-g_L|$ results in more sinusoidal tuning.
For the parameters used in figures of this paper, change in $\omega_{\textrm{ac}}^j$ with flux is less than/equal to $g_{\textrm{eff}}$.

\par
The externally applied flux ($\Phi_e$) preferentially drops across the junction $E_{12}$ which has the highest impedance ($E_{12}\ll E_1,\,E_2$): $E_{12}$ therefore sets coupler hybridization via $g_L(\Phi_e)$. The shunt junctions $E_1$ and $E_2$ protect the coupler hybridization from charge noise, interaction with qubits, and other external circuitry. As a consequence, the decoupling flux bias, $\Phi_e^{(0)}$ s.t. $g_C-g_L(\Phi_e^{(0)})\propto g_{\textrm{eff}}(\Phi_e^{(0)}) = 0$, is \textit{internally defined} by the choice of coupler junction parameters. Figure~\ref{fig:effectivecoupling} shows $2g_{\textrm{eff}}$ as a function of flux $\Phi_e$, for two choices of coupler transmon transitions, above (orange) or below the qubit (blue) transitions. \cc{The shaded region shows the gap between the qubit eigenfrequencies, numerically calculated using Eq.~(\ref{eq:HOfullfull}), where qubit~$b$ is swept into degeneracy with qubit~$a$. These calculations were repeated thirty times with 6\% Gaussian random variation applied independently to the three DTC junction parameters: the width of the shading shows the standard deviation in $2g_{\textrm{eff}}$ at each $\Phi_e$.} Note the insensitivity of $\Phi_e^{(0)}$ to variation in the junction parameters. By contrast, increasing $g_C$ shifts $\Phi_e^{(0)}$ further away from $\Phi_{e}/\Phi_0=0.5$ as shown by comparing Fig.~\ref{fig:effectivecoupling}a-b. \cc{The analytical expression for $g_{\textrm{eff}}$ [Eq.~(\ref{app:effg w anh})] matches numerical simulations to within several percent in the dispersive limit. Equation~(\ref{app:effg w anh}) contains additional terms that correspond to an approximately 20\% correction to Eq.~(\ref{eq:corot}).}

\begin{figure}[t!]
\begin{centering}
\includegraphics[width=\columnwidth]{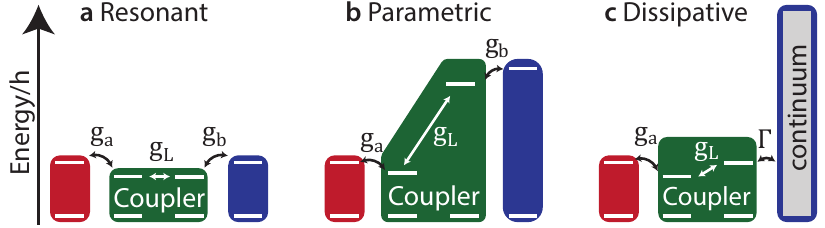}
\caption{\label{fig:usecases}
Quantum state and coupling primitives for resonant, parametric, and dissipative use cases. \textbf{a)} A static coupling between nearly resonant qubits may be turned on with a static $g_L$.
\textbf{b)} Parametric coupling between circuit elements with very different transition frequencies may be accomplished by modulating $g_L$ at the frequency difference. The hybridization of coupler transmons modulates even when it's transmon plasma frequencies differ. Coupling magnitude scales inversely with the relative detuning of each coupler transmon from its qubit.
\textbf{c)} Purcell decay of a qubit through the coupler is turned off when $g_L=0$. This functionality allows waveform shaping of qubit emission into a waveguide.}
\end{centering}
\vspace{-10pt}
\end{figure}
\par
\subsection{Coupling nearly degenerate qubits}

The independence of $\Phi_e^{(0)}$ from the qubit plasma frequency allows for straightforward implementation of degenerate coupling strategies. As an example, consider two qubits, each dispersively coupled to a separate coupler transmon, as shown in Figs.~\ref{fig:usecases}a-b. To implement a controlled-Z operation, data qubit~b is initialized at a frequency $\omega_b$ that is 500~MHz higher than $\omega_a$; having some frequency detuning between the qubits helps to mitigate single-qubit microwave crosstalk. Meanwhile, the DTC flux bias is set to $\Phi_e = \Phi_e^{(0)}$. Next, by tuning the frequency of either qubit, the state $|2000\rangle$ is swept into degeneracy with $|1001\rangle$ (Fig. 3a), where the kets label transmon modes in Fig.~\ref{fig:couplerdispersion}b from left to right. When the states approach degeneracy, $\Phi_e$ is ramped away from $\Phi_e^{(0)}$, turning on a static coupling. The system then evolves for the requisite dwell time at the resulting avoided level crossing. Subsequently, the inverse of the previous flux ramps is implemented to complete the controlled-Z operation~\citep{Chen2014, Yan2018, Sung2021, Xu2020, Collodo2020}. Unlike interference couplers~\citep{Yan2018}, $\Phi_e^{(0)}$ remains a constant throughout the procedure, enhancing ease of use. A procedure for using a DTC to implement controlled-Z between non-degenerate fixed-frequency qubits is also explored in Refs.~\cite{Goto2022, Kubo2022}.
\par
In a similar manner, it is also possible to implement iSWAP operations via standard approaches~\citep{Krantz_apr2019} utilizing the DTC SQuID. It should be noted, though, that even for evolution under the effective Hamiltonian in Eq.~(\ref{eq:simpleH}) -- as is the case for the DTC -- turning on $g_{\textrm{eff}}$ to perform iSWAP between $|1000\rangle$ and $|0001\rangle$ likewise turns on parasitic $\sigma_z\sigma_z$ phase evolution from interactions in the second excitation manifold $|0002\rangle \leftrightarrow |1001\rangle$, which must be separately cancelled~\citep{Sung2021, Mundada2019, Kandala2020, Jaseung2020}.
\par
The DTC shares features with other three island transmon circuits~\citep{Gambetta2011} as well as the capacitively shunted flux (CSFQ) qubit~\cite{Yan2016}. Like the CSFQ, the anharmonicity of DTC transitions change as $\Phi_e$ is swept from zero flux to $0.5\,\Phi_0$. In the  $E_{12}\ll E_1,\,E_2$ regime, the anharmonicity of the symmetric eigenstate transition of two strongly hybridized transmons $\omega_1\approxeq\omega_2$ changes from weakly negative (transmon regime) to near zero. In Fig.~\ref{fig:couplerdispersion}c, the eigenenergy of the symmetric eigenstate can be seen to vary with $\Phi_e$, while the small negative anharmonicity of the antisymmetric eigenstate remains constant with $\Phi_e$. The DTC can eliminate both $\sigma_x\sigma_x$ and $\sigma_z\sigma_z$ interactions at the cancellation flux $\Phi_e = \Phi_e^{(0)}$, irrespective of qubit-qubit detuning. While such cancellations can be achieved in interference based couplers, this necessitates parameter fine-tuning and frequency crowding~\citep{Goto2022}.
\begin{figure}[t!]
\begin{centering}
\includegraphics[width=\columnwidth]{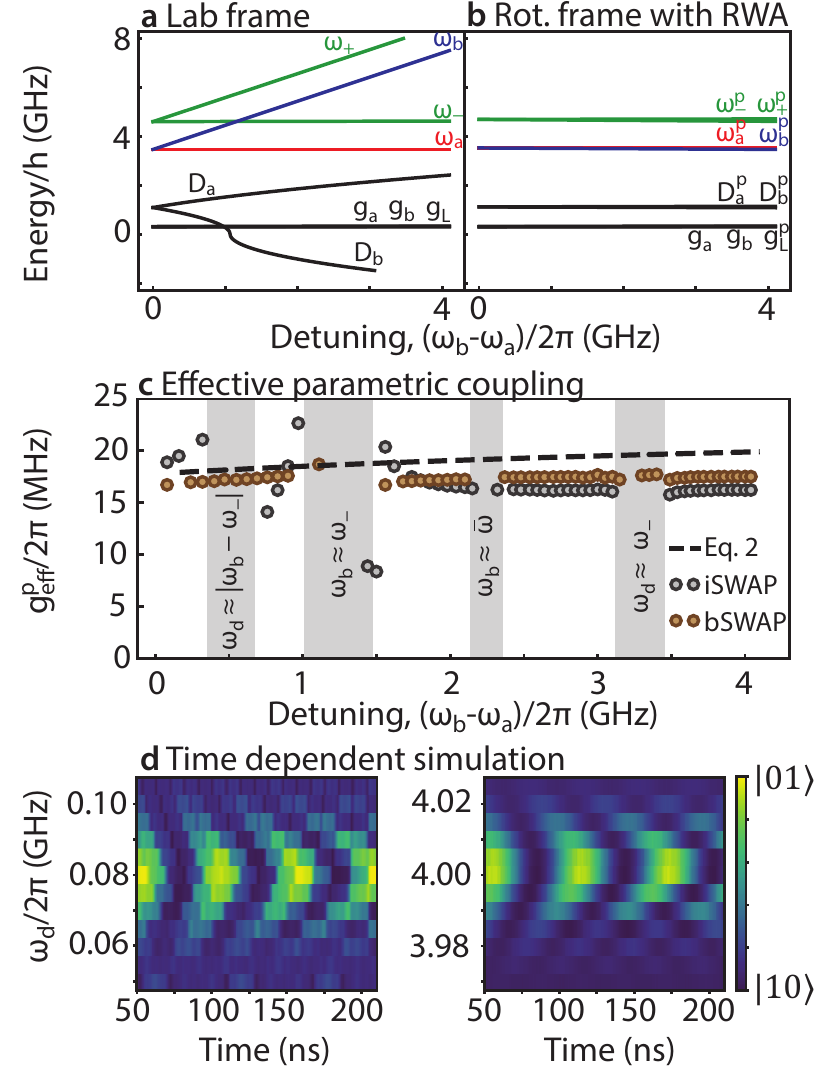}
\caption{\label{fig:parametric_coupling} Model parameters as a function of data transmon detuning defined in the \textbf{a}) lab frame and \textbf{b}) frame rotating with the parametric drive at $\omega_d = \omega_b-\omega_a$, as defined in the main text. The square root denominator $D_j = \textrm{sign}(D_j^2)\sqrt{D_j^2}$ competes with the magnitude of $g_a,\,g_b,\,g_C,\,\textrm{ and }g_L$ to set the size of $g^p_{\textrm{eff}}$. \textbf{c)} The lab frame parameters are chosen such that the rotating frame parameters are constant as a function of data transmon detuning: $g^p_{\textrm{eff}}$ [Eq.~(\ref{app:parametric})] then also remains constant for iSWAP- and bSWAP-style exchange interactions. This picture breaks down in the shaded regions where accidental degeneracies arise between the data and coupler transmon transitions. \cc{The circular markers are resonant exchange rates obtained from 2D fits to the iSWAP or bSWAP time dependent Schr\"{o}dinger equation (TDSE) simulations [Eq.~(\ref{eq:HOfullfull})] performed for different choices of parametric drive frequency $\omega_d/2\pi$ at fixed qubit-qubit detunings $(\omega_b -\omega_a)/2\pi$. \textbf{d)} TDSE iSWAP simulations with $(\omega_b -\omega_a)/2\pi$ equal to $0.08\text{ GHz}$ and $4.00\text{ GHz}$ are shown} in the (left) and (right) panels, respectively. The drive amplitude increases to its maximum value over $50\text{ ns}$ (not shown). The left panel shows substantial contributions from counter-rotating terms.}
\end{centering}
\vspace{-10pt}
\end{figure} 
\par
\section{Coupling with a parametric drive}

We now discuss how the DTC can be employed to implement time-dependent parametric coupling. To this end, a parametric modulation of external flux $\Phi_e = (A \sin{(\omega_d t)}+0.25)\Phi_0$ can mediate Rabi-like exchange interactions with rate $g^p_{\textrm{eff}}$ between fixed-frequency qubits (or between other circuit elements) by tuning $\omega_{d}$ at resonance with the difference of relevant transition frequencies (Fig. 3b).  
The magnitude of resultant parametric coupling $g^p_{\textrm{eff}}$ may be understood in a rotating frame, obtained via the transformation $(a_a^p,\,a_1^p,\,a_b^p,\,a_2^p) = (a_a,\,a_1,\,a_b e^{i\omega_dt},\,a_2 e^{i\omega_dt})$, where qubits $a$ and $b$ are rendered degenerate by the parametric driving [Fig.~\ref{fig:parametric_coupling}(a)]. Neglecting the fast rotating terms, the resulting stationary Hamiltonian is manifestly similar to Eq.~(\ref{eq:H1}) with effective coupling given by Eq.~(\ref{eq:corot}). Substituting $\omega_2^{p} = \omega_2-\omega_d$, $\omega_b^{p} = \omega_b-\omega_d$, and $\bar{\omega}^{p} = \bar{\omega}-\omega_d/2$ into $\Delta_j^{p} = \omega_j^{p}-\bar{\omega}^{p}$ and $2\delta_j^{p} = \omega_1^{p}-\omega_2^{p}$ with $g_L^{p} = (2\pi A-(2\pi A)^3/8)g_L(\Phi_e=0)$, we obtain the parameters given in Fig.~\ref{fig:parametric_coupling}(b) and the effective parametric coupling strength as,
\begin{align}
    g^p_{\textrm{eff}} \approx \sum_{j=a,b} \frac{-g_a g_b g_L^p(1-\Delta^p_j/\bar{\omega}^p)}{4(D_j^p)^2},\label{eq:parametric}
\end{align}
where $ (D_j^p)^{2} = (\Delta_j^{p})^2 - (\delta^p)^{2} - (g_L^p)^2/2 - g_C^2$. While the qubit detuning and pump frequency need to be swept in tandem to maintain resonance $\omega_{d}= \omega_b-\omega_a$, the relative detuning between coupler transmon/qubit can be kept fixed, i.e. $|\omega_a-\omega_1|\sim |\omega_b-\omega_2|$. As a consequence, $\Delta_j^p$ and $\delta_j^p$ also remain constant in the rotating frame as plotted in Fig.~\ref{fig:parametric_coupling}b. Adjusting $g_a,\,g_b,\,\text{and}\,g_L^p$ to remain constant as other variables change [appendix~\ref{app:circ}], $g^p_{\textrm{eff}}$ can be held constant as a function of flux. Under these circumstances numerical simulation confirm a constant $g^p_{\textrm{eff}}$, as shown in Fig.~\ref{fig:parametric_coupling}b, in qualitative agreement with Eq.~(\ref{eq:parametric}). While numerical simulations have included terms up to sixth order in the raising and lowering operators [Fig.~\ref{fig:HOsim}], the analytical expression in Eq.~(\ref{eq:parametric}) is derived using a simplified harmonic oscillator model [Eq.~(\ref{eq:simpleH})] that neglects anharmonicity. However, the expression that generated the dotted line in Fig.~\ref{fig:parametric_coupling}c incorporates small corrections for the anharmonicity [Eq.~(\ref{app:effg w anh})]. When the coupling is turned on, numerical simulations show $\sqrt{2}$ stronger coupling within the second excitation manifold compared to within the first excitation manifold, qualitatively consistent with the simplified analytical model. The qualitative agreement between this model and numerical simulation implies that the anharmonicity of the coupler transmons does not play a strong role in setting $g_{\textrm{eff}}$ or $g^p_{\textrm{eff}}$. 
\par
Another consideration for parametric coupling is the rectification of data qubit frequency due to flux modulation. Specifically, $\omega_{\textrm{ac}}^j$ of the driven qubits scales inversely with $1/(\omega_j-\omega_{\pm})$, which can change nonlinearly with flux; if the parametric drive periodically brings $\omega_j$ close to $\omega_{\pm}$, it can introduce a time-averaged drive amplitude dependence to the Rabi resonance condition that manifests as distortion of the time-domain swap envelope. 
As shown by the simulations presented in Fig.~\ref{fig:parametric_coupling}c, such distortion remains negligible if the coupler-induced time-averaged frequency shifts are less than $0.1\times g^p_{\textrm{eff}}$. 
Relative to interference-based couplers, the DTC mitigates the deleterious impact of the nonlinear flux dependence shifts in two ways. First, the DTC eigenenergies tune gently with effective coupling. Second, the freedom to set the detuning and interaction rate $g_j$ separately for each DTC transmon/qubit pair in fabrication allows the tuning of $\omega_{\textrm{ac}}^a$ and $\omega_{\textrm{ac}}^b$ with flux to be balanced. This balancing strategy roughly maximizes $g^p_{\textrm{eff}}$ for a fixed magnitude of nonlinearity. The combination of these factors allows straightforward, and relatively fast, pairwise parametric coupling between non-degenerate fixed-frequency qubits, among other applications. The parameters used in Figs.~\ref{fig:effectivecoupling} and \ref{fig:parametric_coupling} represent a worst case scenario for second order nonlinearity generation, where $\delta \sim 0$. Even so, parametric coupling of qubits with $80$~MHz relative detuning showed Rabi-like state evolution, such as that shown in Fig.~\ref{fig:parametric_coupling}d, persists as $E_{12}$ increases $10\%$~($1\%$) for the coupler transitions placed below~(above) the qubit transitions. 
\par
A parametric drive can also induce unwanted direct energy exchange between various states in the Hilbert space, i.e. $\omega_d\approx |\omega_j-\omega_{\pm}|$, that should be avoided through appropriate choice of coupler transitions in fabrication. This is an important consideration for small qubit detuning. For larger qubit detunings, some driving frequencies will excite the DTC directly $\omega_d\sim \omega_{\pm}$. A weak, undesirable, multi-photon exchange interactions will also occur in the second excited manifold $\ket{1001}\leftrightarrow \ket{0110}$ when $\bar{\omega} = (\omega_a+\omega_b)/2$: choosing the level structure as depicted in Figs.~\ref{fig:parametric_coupling}~and~\ref{fig:usecases}a-b avoids this undesirable degeneracy.
\par
\section{Coupling to a waveguide}
The last use case that we consider is fast tunable coupling between a waveguide (or other open quantum system) and a qubit, as depicted in Fig.~\ref{fig:usecases}c. This use case is thematically consistent with previous demonstrations~\citep{Besse2020, Kannan2022}. For this scenario we make the left data qubit and the right DTC transmon degenerate, i.e. $\omega_a = \omega_2$, which invalidates the dispersive approximation used in Eq.~(\ref{eq:simpleH}). Instead, the effective tunable coupling is given by $g_a(g_C-g_L)/2\delta$, which goes to zero when the flux is set to $\Phi_e^{(0)}$. If the degeneracy between the qubit and DTC transmon is broken, the qubit relaxation rate into the waveguide under the dispserive limit is given by $\Gamma g_a^2(g_C-g_L)^2/D_a^4$, which is typically small. Further, for $\omega_a\ne\omega_2$, the use cases described in Figs.~\ref{fig:usecases}b-c can be combined to realize strong coupling to a continuum without demanding degeneracy between qubit and DTC transitions.

\section{Energy relaxation and coherence considerations}
Our DTC layout (Fig.~\ref{fig:couplerdispersion}a) uses a grounded coplanar waveguide with impedance $Z_0 = 50 \text{ }\Omega$ that is shorted near the three-junction SQuID loop: current in the waveguide induces a controllable flux $\Phi_e$ via a mutual inductance $M$ to the loop. The three junction DTC SQuID allows direct exchange coupling between the DTC states and current excitations in the flux bias line. Using the circuit model discussed in appendix~\ref{app:noise}, the resultant uncorrelated relaxation rate of each DTC transmon into the bias line can be approximated as, $\Gamma_{1,j} \approx  M^2E_{12}^2\cos^2{(\Phi_e/\varphi_0)}/Z_0\varphi_0^4C_j$ for $j\in \{1,2\}$, which only doubles if $g_C$ is increased from zero to $g_C\sim g_L$: see the expression for $\xi_j$ and the derivation for $\Gamma_{1,j}$ in the appendix~\ref{app:noise}. Therefore, DTCs with small mutual inductive coupling $M\leq 3\text{ pH}$ likely will not need additional low pass filters on their current bias lines so long as the thermal population of the flux bias line is low. Conveniently, $\Gamma_1$ also tunes to zero at $g_L = 0$: thus, energy relaxation of the coupler into its flux bias line occurs only when $g_{\textrm{eff}} \ne 0$ in the $g_C \ll g_L$ limit. 
\par
Spontaneous emission of the DTC, whatever the underlying source of fluctuations, is a source of Purcell decay for the qubits whose rate can be approximated as, $\Gamma_{1,a} = (g_a/\Delta_a)^2\Gamma_{1,1}$ and $\Gamma_{1,b} = (g_b/\Delta_b)^2\Gamma_{1,2}$ where $\Gamma_{1,j}$ for $j\in \{a,1,2,b\}$ is the relaxation rate of the relevant transmon. Multi-junction transmon devices routinely have lifetimes in excess of tens of microseconds. Assuming $T_1 > 20\text{ }\mu\text{s}$ and a Purcell factor $(g_j/\Delta_j)^2\sim 1/10$, the coupler transmons impose at worst a $200\text{ }\mu\text{s}$ energy relaxation time on each qubit. While the dispersive regime is considered to be $(g_j/\Delta_j) < 0.1$, violating the dispersive approximation condition requires renormalization of $(g_j/\Delta_j)$ and the inclusion of fourth order terms in the Schrieffer-Wolff transformation for quantitative accuracy~\citep{Xiao2021} and accounts for the deviation between Eqs.~(\ref{app:effg w anh}) and (\ref{app:parametric}) and numerical simulations in Figs.~\ref{fig:effectivecoupling}~and~\ref{fig:parametric_coupling}, respectively.
\par
The impact of DTC-induced dephasing on the qubits can be computed from the frequency response of its eigenstates in presence of fluctuations on loop flux of the SQuID. The coupler eigenenergies change by at most $4g_L$ over the full flux tuning range, with a maximum slope at the idling flux typically set to the deocupling bias $\Phi_{e}^{(0)}$. At the idling flux, the slope is then $|\partial\omega/\partial\Phi| = 2g_L/\varphi_0$. For typical amplitudes of flux noise spectral density, $\sqrt{A_{\Phi}} \sim 2.5\mu\Phi_0/\sqrt{\textrm{Hz}}$, the Hahn echo dephasing rate due to $1/f$ flux noise can be estimated as $\Gamma^{E}_{\phi} = \sqrt{A_{\phi}\textrm{ln}(2)}|\partial\omega/\partial\Phi| \sim 50\times 10^3\text{ s}^{-1/2}$~\cite{Braumuller2020}. The frequency fluctuations of the qubits due to flux- and critical-current noise of the coupler transmons are filtered by the Purcell factor $(g_j/\Delta_j)^2$. Hence for a Purcell factor of $1/10$, $1/f$ flux noise would limit the coherence of coupled qubit modes to $\sim 200\text{ }\mu\text{s}$ at $\Phi_{e}^{(0)}$.
\par
In summary, we have theoretically explored the use of hybridization between two transmons, a DTC, as a mechanism for mediating tunable interactions between fixed-frequency qubits. The DTC design offers low noise, along with linear and easy-to-model qubit-qubit interactions. Crucially, the decoupling condition is defined by a flux bias that nulls the hybridization, making the decoupling condition almost entirely independent of the properties of the data qubit and qubit-DTC interaction. The extra design freedom afforded by having two coupler transmons rather than one also enables implementation of fast, parametric coupling to a fixed mode or continuum. 
Moreover, the DTC may be attractive as a standalone qubit for its gentle eigenfrequency tuning with flux, weak exchange coupling with its current bias line, and level structure. The last benefit may also enable novel qutrit-based superconducting circuit architectures~\cite{kiktenko2020scalable,baker2020efficient}.

\begin{acknowledgments}
This material is based upon work supported by the U.S. Department of Energy, Office of Science, under award number DE-SC0019461, and Air Force Office of Scientific Research (AFOSR) under grant FA9550-21-1-0151. DLC and ML would also like to acknowledge independent support from AFOSR. Any opinions, findings, and conclusions or recommendations expressed in this article are those of the authors and do not necessarily reflect the views of the Air Force Research Laboratory (AFRL).
\end{acknowledgments}
\appendix
\section{Circuit derivation for the coupler alone}
\label{app:circ}
\begin{figure}
\begin{centering}
\includegraphics[width=3.0in]{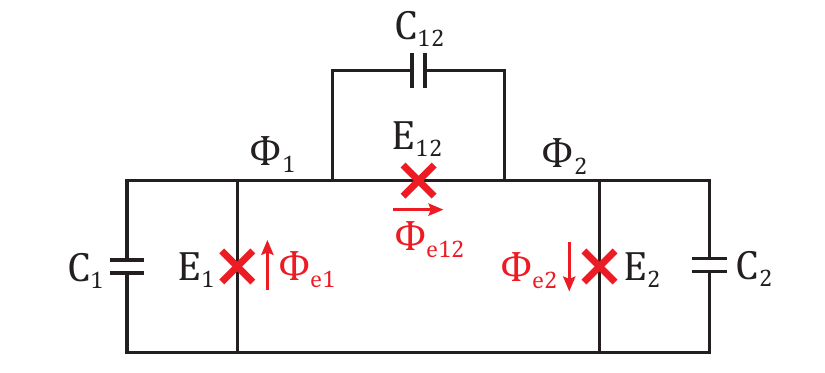}
\caption{\label{fig:TwoIslandCircuit} Coupler circuit comprising two transmons qubits arranged in a flux-tunable loop with a third junction $(E_{12} ,C_{12})$.}
\end{centering}
\end{figure} 

We write down the Lagrangian for the circuit shown in Fig.~\ref{fig:TwoIslandCircuit} using standard circuit parametrization~\cite{riwar2021circuit, Vool_2017, You2019}
\begin{align}
\mathcal{L} = &\frac{1}{2}C_1 (\dot\Phi_1+\dot\Phi_{e1})^2+\frac{1}{2}C_{12}(\dot\Phi_2-\dot\Phi_1+\dot\Phi_{e12})^2\nonumber\\
&\, +\frac{1}{2}C_2 (\dot\Phi_2-\dot\Phi_{e2})^2\nonumber\\
&\, +E_1\cos{((\Phi_1+\Phi_{e1})/\varphi_0)}+E_2\cos{((\Phi_2-\Phi_{e2})/\varphi_0)}\nonumber\\
&\, +E_{12}\cos{((\Phi_2-\Phi_1+\Phi_{e12})/\varphi_0)}\label{eq:Linit},
\end{align}
where we use $\varphi_0$ to denote the reduced flux quantum $\Phi_0/2\pi$. Here $\Phi_j(t)$ represent the node fluxes (a ``position"-like variable) and $\dot\Phi_j(t)$ represent the node voltages (associated ``velocities") respectively. The loop of three Josephson junctions is threaded by an external flux, ${\Phi_e = \Phi_{e12}+\Phi_{e1}+\Phi_{e2}}$, with parameters $\{\Phi_{e12},\,\Phi_{e1},\,\Phi_{e2}\}$ representing the external flux drops across the three junctions $\{E_{12},\,E_1,\,E_2\}$ respectively.
%

Performing the Legendre transformation, we arrive at the Hamiltonian,
\begin{align}
H = &\frac{C_{12}(Q_1+Q_2)^2}{2C^2}+\frac{C_2Q_1^2+C_1Q_2^2}{2C^2}\nonumber\\
&\, +\frac{\dot\Phi_{e12} (-Q_1C_2C_{12}+Q_2C_1C_{12})}{C^2}\nonumber\\
&\, +\frac{\dot\Phi_{e1} (Q_1(C_1C_2+C_1C_{12})+Q_2C_1C_{12})}{C^2}\nonumber\\
&\, -\frac{\dot\Phi_{e2} (Q_1C_1C_{12}+Q_2(C_1C_2+C_1C_{12}))}{C^2}\nonumber\\
&\, -E_1\cos{((\Phi_1+\Phi_{e1})/\varphi_0)}-E_2\cos{((\Phi_2- \Phi_{e2})/\varphi_0)}\nonumber\\
&\, -E_{12}\cos{((\Phi_2-\Phi_1+\Phi_{e12})/\varphi_0)},\label{eq:trueH}
\end{align}
where $C^2 = C_1C_2+C_1C_{12}+C_2C_{12}$. Using the loop constraint we may parametrize $\{\Phi_{e12},\,\Phi_{e1},\,\Phi_{e2}\}$ in terms of $\Phi_e$ such that all $\dot\Phi_e$ terms cancel~\cite{You2019}:
\begin{align}
H = &\frac{C_{12}(Q_1+Q_2)^2}{2C^2}+\frac{C_2Q_1^2+C_1Q_2^2}{2C^2}\nonumber\\
&\,-E_1\cos{\left[\left(\Phi_1+\frac{C_{12}C_2}{C^2}\Phi_e\right)/\varphi_0\right]}\nonumber\\
&\,-E_2\cos{\left[\left(\Phi_2-\frac{C_{12}C_1}{C^2}\Phi_e\right)/\varphi_0\right]}\nonumber\\
&\,-E_{12}\cos{\left[\left(\Phi_2-\Phi_1+\frac{C_1C_2}{C^2}\Phi_e\right)/\varphi_0\right]}\label{eq:chargeH}.
\end{align}
In this parameterization, we can infer that for typical capacitive coupling, $C_{12}\leq 0.1 \sqrt{C_1C_2}$, the participation of $E_{12}$ with $\Phi_e$ is greater than $80\%$. Due to the lack of $\dot\Phi_e$ terms, Eq.~(\ref{eq:chargeH}) is the simplest construction for charge basis simulations of the Hamiltonian.

Another useful parameterization redistributes $\Phi_e$ across the junctions to set $\partial H/\partial \Phi_j = 0$ for $j=\{1,\,2\}$. This cancels odd order terms in $\Phi_j$ and thereby enables a straightforward conversion to the harmonic oscillator basis. We assume $\langle \Phi_j^2 \rangle \ll \Phi_0$ and expand to fourth order in $\Phi_j$
\begin{align}
H \approx &\frac{C_{12}(Q_1+Q_2)^2}{2C^2}+\frac{C_2Q_1^2+C_1Q_2^2}{2C^2}\nonumber\\
&\, +\frac{\dot\Phi_{e12} (-Q_1C_2C_{12}+Q_2C_1C_{12})}{C^2}\nonumber\\
&\, +\frac{E'_{12}\dot\Phi_{e12}}{E'_{1}}\frac{(Q_1(C_1C_2+C_1C_{12})+Q_2C_1C_{12})}{C^2}\nonumber\\
&\, -\frac{E'_{12}\dot\Phi_{e12}}{E'_{2}}\frac{(Q_1C_2C_{12}+Q_2(C_1C_2+C_2C_{12}))}{C^2}\nonumber\\
&\, +E'_{1}\left(\frac{\Phi_1^2}{2\varphi_0^2}-\frac{\Phi_1^4}{24\varphi_0^4}\right)\nonumber\\
&\, +E'_{2}\left(\frac{\Phi_2^2}{2\varphi_0^2} - \frac{\Phi_2^4}{24\varphi_0^4}\right)\nonumber\\
&\, +E'_{12}\left(\frac{(\Phi_2-\Phi_1)^2}{2\varphi_0^2} - \frac{(\Phi_2-\Phi_1)^4}{24\varphi_0^4}\right)\label{eq:preharmonicH}.
\end{align}
The external flux is then 
\begin{align}
\Phi_e = \Phi_{e12} + &\varphi_0\arcsin{((E_{12}/E_1)\sin{(\Phi_{e12}/\varphi_0)})}\nonumber\\+&\varphi_0\arcsin{((E_{12}/E_2)\sin{(\Phi_{e12}/\varphi_0)})},
\end{align}
which may be numerically inverted. The junction energies are now flux sensitive: $E'_{12} = E_{12}\cos{(\Phi_{e12}/\varphi_0)}$ and $E'_{j} = \sqrt{E_j^2-E_{12}^2\sin^2{(\Phi_{e12}/\varphi_0)}}$.

\subsection{Harmonic oscillator basis}

\begin{figure}
\begin{centering}
\includegraphics[width=3.3in]{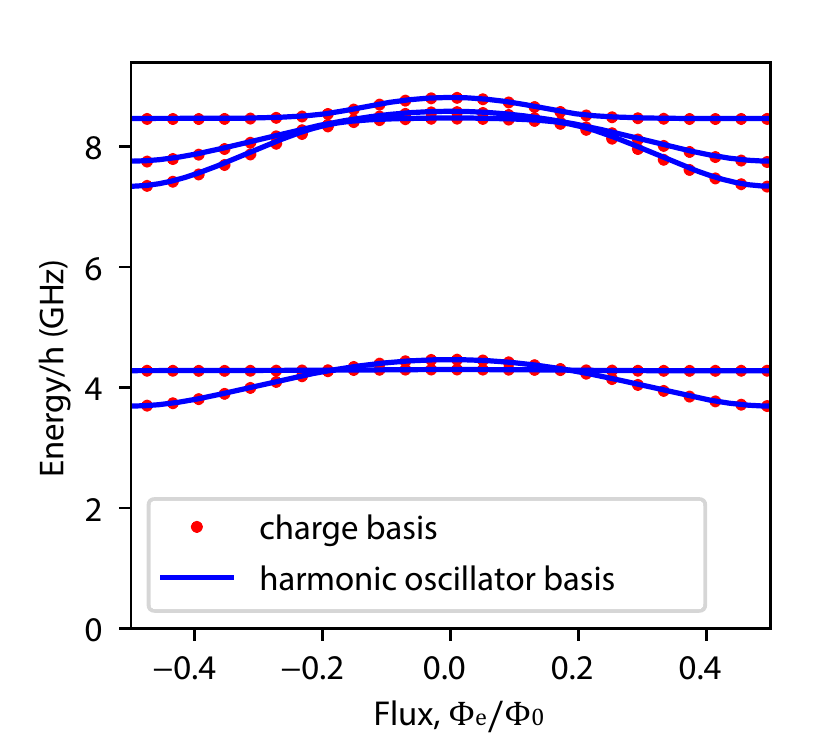}
\caption{Comparison of the simulated eigenenergies found by a harmonic oscillator basis simulation [Eq.~(\ref{eq:chargeH})] verses a charge basis simulation [\ref{eq:HOfull}]. We chose parameters $E_{j}=13\textrm{ GHz}$, $E_{12} = 1.3\textrm{ GHz}$, $C_j=100\textrm{ fF}$, and $C_{12} = 0\textrm{ fF}$. We simulated 8 levels per qubit for the harmonic oscillator simulation and 14 Cooper pair levels in the charge simulation.}
\end{centering}
\label{fig:HOsim}
\end{figure}

We convert to the Harmonic oscillator basis by substituting ${Q_j = \sqrt{\hbar\omega_jC_{Sj}/2}(a_j + a_j^{\dagger})}$ and ${\Phi_j = -i\sqrt{\hbar\omega_jL_j/2}(a_j - a_j^{\dagger})}$ in Eq.~(\ref{eq:preharmonicH}). This leads to the following approximate Hamiltonian, with parameters defined in Table~\ref{table: parameters}, as
\begin{table}
\begin{centering}
\begin{tabular}{c|c}
    term & value\\
\hline
    $C_{S1}$ & $\frac{C^2}{C_{12}+C_2}$\\
    $C_{S2}$ & $\frac{C^2}{C_{12}+C_1}$\\
    $L_j$ & $\frac{\varphi_0^2}{E'_{12}+E'_{j}}$ \\
    $E_{Cj}$ & $e^2/2C_{Sj}$\\
    $g_C$ & $\frac{C_{12}\sqrt{\omega_1\omega_2}}{2 \sqrt{(C_1+C_{12})(C_2+C_{12})}}$\\
    $g_L$ & $\frac{4E'_{12}\sqrt{E_{C1}E_{C2}}}{\hbar^2\sqrt{\omega_1\omega_2}}$\\
    $\xi_1$ & $\frac{(2(E_1'+E_{12}'))^{1/4}[-E_1'(E_2'+E_{12}')C_2C_{12}+E_2'E_{12}'(C_1C_2+C_1C_{12})]}{2E_{C2}^{1/4}E_1'E_2'C^2}$\\
    $\xi_2$ & $\frac{(2(E_2'+E_{12}'))^{1/4}[E_2'(E_1'+E_{12}')C_1C_{12}-E_1'E_{12}'(C_1C_2+C_2C_{12})]}{2E_{C2}^{1/4}E_1'E_2'C^2}$\\
    $\omega_j$ & $\frac{\sqrt{8(E_j'+E_{12}')E_{Cj}}}{\hbar}$\\
    $\nu_j^{(4)}$ & $\frac{E_{Cj}}{12\hbar}$\\
    $\nu_j^{(6)}$ & $\frac{\sqrt{2}E_{Cj}}{360\hbar}\left(\frac{E_{Cj}}{E_j'+E_{12}'}\right)^{1/2}$\\
    $\mu_k^{(4)}$ & $\binom{4}{k}\frac{(-1)^kE'_{12}}{12\hbar}\left(\frac{E_{C1}}{E_1'+E_{12}'}\right)^{k/4}\left(\frac{E_{C2}}{E_2'+E_{12}'}\right)^{(4-k)/4}$\\
    $\mu_k^{(6)}$ & $\binom{6}{k}\frac{(-1)^k\sqrt{2}E'_{12}}{360\hbar}\left(\frac{E_{C1}}{E_1'+E_{12}'}\right)^{k/4}\left(\frac{E_{C2}}{E_2'+E_{12}'}\right)^{(6-k)/4}$
\end{tabular}
\caption{\label{table: parameters} Parameter definitions}
\end{centering}
\end{table}
%
%
%
\begin{align}
    H/\hbar =& \sum_{j=1}^2\left[\omega_ja_j^{\dagger}a_j-\nu_j^{(4)}(a_j-a_j^{\dagger})^4-\nu_j^{(6)}(a_j-a_j^{\dagger})^6\right]\nonumber\\
    &\, +g_C(a_1+a_1^{\dagger})(a_2+a_2^{\dagger})+g_L(a_1-a_1^{\dagger})(a_2-a_2^{\dagger})\nonumber\\
    &\, +\sum_{j=1}^{2}\frac{\xi_j\dot\Phi_{e12}}{\Phi_0}(a_j+a_j^{\dagger})\nonumber\\
    &\, -\sum_{k=1}^3\mu_k^{(4)}(a_1-a_1^{\dagger})^k(a_2-a_2^{\dagger})^{(4-k)}\nonumber\\
    &\, -\sum_{k=1}^3\mu_k^{(6)}(a_1-a_1^{\dagger})^k(a_2-a_2^{\dagger})^{(6-k)}
    \label{eq:HOfull}.
\end{align}
The terms fourth order in the raising and lowering operators modify the eigenenergies in the harmonic oscillator basis by several hundred MHz and account for the majority of the anharmonicity. The sixth order terms compensate for a residual few 10s of MHz of discrepancy between the charge basis and harmonic oscillator approaches to simulation. At sixth order in the raising and lowering operators the first and second excitation manifolds of eigenenergies for the two simulation approaches agree to within 1~MHz, shown together in Fig.~\ref{fig:HOsim}.
%

\section{Circuit derivation for coupler and data qubits}
\label{app:full}

We now include the data qubit terms to the Lagrangian in Eq.~(\ref{eq:Linit}), collect the capacitive terms between operators, and break out the time dependence of the flux
\begin{align}
\mathcal{L} = &\frac{1}{2}\vec{\Phi}^T \check{C}\vec{\Phi}\nonumber\\
&\, +\dot\Phi_1(C_1\dot\Phi_{e1}-C_{12}\dot\Phi_{e12})-\dot\Phi_2(C_2\dot\Phi_{e2}-C_{12}\dot\Phi_{e12})\nonumber\\
&\, +E_1\cos{((\Phi_1+\Phi_{e1})/\varphi_0)}+E_2\cos{((\Phi_2-\Phi_{e2})/\varphi_0)}\nonumber\\
&\, +E_{12}\cos{((\Phi_2-\Phi_1+\Phi_{e12})/\varphi_0)}\nonumber\\
&\, +E_a\cos{(\Phi_a/\varphi_0)}+E_b\cos{(\Phi_b/\varphi_0)},
\end{align}
where $\vec{\Phi} = (\dot\Phi_a,\,\dot\Phi_1,\,\dot\Phi_2,\,\dot\Phi_b)^T$ and the capacitance matrix is given by,
\begin{align}
\begin{pmatrix}
C_a & C_{a1} & 0 & 0\\
C_{a1} & C_1 & C_{12} & 0\\
0 & C_{12} & C_2 & C_{b2}\\
0 & 0 & C_{b2} & C_b
\end{pmatrix}.
\end{align}

Performing the Legendre transformation, we arrive at the Hamiltonian,
\begin{align}
\mathcal{H} = &\frac{1}{2}\vec{Q}^T \check{C}^{-1}\vec{Q}\nonumber\\
&\, + \vec{Q}^{T}\check{C}^{-1}
\begin{pmatrix}
0\\
C_1\dot\Phi_{e1}-C_{12}\dot\Phi_{e12}\\
-C_2\dot\Phi_{e2}+C_{12}\dot\Phi_{e12}\\
0
\end{pmatrix}\nonumber\\
&\, +E_1\cos{((\Phi_1+\Phi_{e1})/\varphi_0)}+E_2\cos{((\Phi_2-\Phi_{e2})/\varphi_0)}\nonumber\\
&\, +E_{12}\cos{((\Phi_2-\Phi_1+\Phi_{e12})/\varphi_0)}\nonumber\\
&\, +E_a\cos{(\Phi_a/\varphi_0)}+E_b\cos{(\Phi_b/\varphi_0)}\label{eq:chargeHfullunparam},
\end{align}
where $\vec{Q}^T = (Q_a,\,Q_1,\,Q_2,\,Q_b)$ and $Q_j$ for $j\in\{a,\,1,\,2,\,b\}$ are charge operators for the $j$'th transmon. If $C_{1a}$, $C_{2b}$, and $C_{12}$ are all non-zero, then direct capacitive coupling exists between all qubits.

Using the loop constraint we may parametrize $\{\Phi_{e12},\,\Phi_{e1},\,\Phi_{e2}\}$ such that all $\dot\Phi_e$ terms cancel~\cite{You2019}:
\begin{align}
H = &\frac{1}{2}\vec{Q}^T \check{C}^{-1}\vec{Q}\nonumber\\
&\,-E_1\cos{\left[\left(\Phi_1+\frac{C_{12}C_2}{C^2}\Phi_e\right)/\varphi_0\right]}\nonumber\\
&\,-E_2\cos{\left[\left(\Phi_2-\frac{C_{12}C_1}{C^2}\Phi_e\right)/\varphi_0\right]}\nonumber\\
&\,-E_{12}\cos{\left[\left(\Phi_2-\Phi_1+\frac{C_1C_2}{C^2}\Phi_e\right)/\varphi_0\right]}\label{eq:chargeHfull}.
\end{align}
Interestingly, this parameterization of the flux does not change from that of the coupler alone in Eq.~(\ref{eq:chargeH}).  Due to the lack of $\dot\Phi_e$ terms, Eq.~(\ref{eq:chargeHfull}) is the simplest construction for charge basis simulations of the full qubit and coupler Hamiltonian.

\subsection{Approximations for harmonic oscillator basis}

Another useful parameterization involves expanding the cosine terms in the Hamiltonian according to 
\begin{align}
\cos(\Phi+\Phi_e)&\approx \cos(\Phi)\cos(\Phi_e)+\sin(\Phi)\sin(\Phi_e)\\
& \approx \cos(\Phi)\cos(\Phi_e)+\Phi \sin(\Phi_e)
\end{align}
where we have assumed $\langle \Phi_j^2 \rangle \ll \Phi_0$, which is appropriate in the transmon limit $E_J/E_C > 50$. 

If, after expanding in $\Phi_j$ and $\dot{\Phi}_j$, the largest terms are proportional to $\Phi_j^2$ and $\dot{\Phi}_j^2$ the harmonic oscillator basis can require fewer states to be modeled relative to the charge basis while achieving equivalent numerical precision. In flux-sensitive systems this can require some extra work since the external flux must be distributed across the junctions so that terms linear in $\Phi_j$ cancel. Solving the following equations, separately for $\Phi_{e1}$ and $\Phi_{e2}$, performs this cancellation:
\begin{align}
E_1 \sin(\Phi_{e1}) &= E_{12} \sin(\Phi_{e12})\\
E_2 \sin(\Phi_{e2}) &= E_{12} \sin(\Phi_{e12})\nonumber\\
\Phi_{e1} &= \arcsin(E_{12}\sin(\Phi_{e12})/E_1)\\
\Phi_{e2} &= \arcsin(E_{12}\sin(\Phi_{e12})/E_2)\nonumber\\
\dot{\Phi}_{e1} &= E_{12}'\dot{\Phi}_{e12}/E_1'\\
\dot{\Phi}_{e2} &= E_{12}'\dot{\Phi}_{e12}/E_2'\nonumber\\
E_{12}' &= E_{12}\cos(\Phi_{e12}/\varphi_0)\label{eq:E12}\\
E_{1}' &= \sqrt{E_1^2-E_{12}^2\sin^2(\Phi_{e12}/\varphi_0)}\label{eq:E1}\\
E_{2}' &= \sqrt{E_2^2-E_{12}^2\sin^2(\Phi_{e12}/\varphi_0)}\label{eq:E2}
\end{align}
Substituting into Eq.~(\ref{eq:chargeHfullunparam}) and truncating at sixth order in $\Phi_j$ we arrive at a suitable starting point for a transformation to the harmonic oscillator basis:
\begin{align}
H \approx &\frac{1}{2}\vec{Q}^T \check{C}^{-1}\vec{Q}\label{eq:preharmonicHfull}\\
&\, + \vec{Q}^{T}\check{C}^{-1}
\begin{pmatrix}
0\\
(C_1E_{12}'/E_1'-C_{12})\dot\Phi_{e12}\\
(-C_2E_{12}'/E_2'+C_{12})\dot\Phi_{e12}\\
0
\end{pmatrix}\nonumber\\
&\, +E'_{1}\left(\frac{\Phi_1^2}{2\varphi_0^2}-\frac{\Phi_1^4}{24\varphi_0^4}+\frac{\Phi_1^6}{720\varphi_0^6}\right)\nonumber\\
&\, +E'_{2}\left(\frac{\Phi_2^2}{2\varphi_0^2} - \frac{\Phi_2^4}{24\varphi_0^4}+\frac{\Phi_2^6}{720\varphi_0^6}\right)\nonumber\\
&\, +E'_{12}\left(\frac{(\Phi_2-\Phi_1)^2}{2\varphi_0^2} - \frac{(\Phi_2-\Phi_1)^4}{24\varphi_0^4} + \frac{(\Phi_2-\Phi_1)^6}{720\varphi_0^6}\right)\nonumber.
\end{align}
$E_1'$, $E_2'$ and $E_{12}'$ are parametrized by $\Phi_{e12}$. $\Phi_{e12}$ is related to the true externally applied flux $\Phi_e$ by the following relation
\begin{align}
\Phi_e = \Phi_{e12} + &\varphi_0\arcsin{((E_{12}/E_1)\sin{(\Phi_{e12}/\varphi_0)})}\nonumber\\+&\varphi_0\arcsin{((E_{12}/E_2)\sin{(\Phi_{e12}/\varphi_0)})},\label{eq:trueflux}
\end{align}
which may be numerically inverted.

\subsection{Hamiltonian for the full circuit in the harmonic oscillator basis}

We convert to the Harmonic oscillator basis by substituting ${Q_j = \sqrt{\hbar\omega_jC_{Sj}/2}(a_j + a_j^{\dagger})}$ and ${\Phi_j = -i\sqrt{\hbar\omega_jL_j/2}(a_j - a_j^{\dagger})}$ in Eq.~(\ref{eq:preharmonicHfull}) for $j\in\{a,\,1,\,2,\,b\}$. 

This leads to the following approximate Hamiltonian given in Eq.~(\ref{eq:HOfullfull}), with parameters defined in Table~\ref{table: parametersfull}
\begin{table}
\begin{centering}
\begin{tabular}{c|c}
    term & value\\
\hline
    $C_{Sj,j}$ & $1/C^{-1}_{j,j}$\\
    $L_j$ for $j\in\{1,\,2\}$ & $\varphi_0^2/(E'_{12}+E'_{j})$ \\
    $L_j$ for $j\in\{a,\,b\}$ & $\varphi_0^2/E_{j}$ \\
    $Z_j$ & $\sqrt{L_j/C_{Sj,j}}$\\
    $E_{Cj}$ & $e^2/2C_{Sj,j}$\\
    $g_{Cj,k}$ & $C^{-1}_{j,k}/2\sqrt{Z_jZ_k}$\\
    $g_L$ & $E_{12}'\sqrt{Z_jZ_k}/2\varphi_0^2$\\
    $\xi_j$ & $\begin{aligned}\Phi_0&\big[C^{-1}_{j,1}(C_1E_{12}'/E_1'-C_{12})\\+&C^{-1}_{j,2}(-C_2E_{12}'/E_2'+C_{12})\big]/\sqrt{2\hbar Z_j}\end{aligned}$\\
    $\hbar\omega_j$ & $\sqrt{8(E_j'+E_{12}')E_{Cj}}$\\
    $\nu_j^{(4)}$ & $\frac{E_{Cj}}{12\hbar}$\\
    $\nu_j^{(6)}$ & $\frac{\sqrt{2}E_{Cj}}{360\hbar}\left(\frac{E_{Cj}}{E_j'+E_{12}'}\right)^{1/2}$\\
    $\mu_k^{(4)}$ & $\binom{4}{k}\frac{(-1)^kE'_{12}}{12\hbar}\left(\frac{E_{C1}}{E_1'+E_{12}'}\right)^{k/4}\left(\frac{E_{C2}}{E_2'+E_{12}'}\right)^{(4-k)/4}$\\
    $\mu_k^{(6)}$ & $\binom{6}{k}\frac{(-1)^k\sqrt{2}E'_{12}}{360\hbar}\left(\frac{E_{C1}}{E_1'+E_{12}'}\right)^{k/4}\left(\frac{E_{C2}}{E_2'+E_{12}'}\right)^{(6-k)/4}$
\end{tabular}
\caption{\label{table: parametersfull} Parameter definitions}
\end{centering}
\end{table}

\begin{align}
    H/\hbar =& \sum_{j}\left[\omega_ja_j^{\dagger}a_j-\nu_j^{(4)}(a_j-a_j^{\dagger})^4-\nu_j^{(6)}(a_j-a_j^{\dagger})^6\right]\nonumber\\
    &\, +\sum_{j \ne k}g_{Cj,k}(a_j+a_j^{\dagger})(a_k+a_k^{\dagger})\nonumber\\
    &\, +g_L(a_1-a_1^{\dagger})(a_2-a_2^{\dagger})\nonumber\\
    &\, +\sum_{j}\frac{\xi_j\dot\Phi_{e12}}{\Phi_0}(a_j+a_j^{\dagger})\nonumber\\
    &\, -\sum_{\ell=1}^3\mu_k^{(4)}(a_1-a_1^{\dagger})^k(a_2-a_2^{\dagger})^{(4-k)}\nonumber\\
    &\, -\sum_{\ell=1}^3\mu_k^{(6)}(a_1-a_1^{\dagger})^k(a_2-a_2^{\dagger})^{(6-k)}
    \label{eq:HOfullfull}
\end{align}
where the summations range over $j,\,k\in \{a,\,1,\,2,\,b\}$.

The inverted capacitance matrix results in effective capacitive coupling between all four transmons in the system. We represent the exact analytic form followed by three approximate examples
\begin{align}
    g_{j,k} &= \frac{|A_{s\backslash j,s \backslash k}|\sqrt{\omega_{j}\omega_{k}}}{2\sqrt{|A_{s\backslash j,s \backslash j}||A_{s\backslash k,s \backslash k}|}}\\
    g_{a,1} &\approx C_{a1}\sqrt{\omega_a\omega_1}/C_S\label{eq:ga1}\\
    g_{a,2} &\approx C_{a1}C_{12}\sqrt{\omega_a\omega_2}/C_S^2\\
    g_{a,b} &\approx C_{a1}C_{b2}C_{12}\sqrt{\omega_a\omega_b}/C_S^3\label{eq:gab}
\end{align}
where $s =\{a,1,2,b\}$, $A$ is the capacitance matrix from the Lagrangian, $|A|$ is the determinant of that matrix, and $A_{s\backslash j,s \backslash k}$ is the four transmon capacitance matrix with row $j$ and column $k$ deleted. In Eqs.~(\ref{eq:ga1}-\ref{eq:gab}), $C_S$ is a stand-in for something with the approximate value of the shunt capacitance, such as $C_{Sj}$. The approximate relations $g_{a,1}$, $g_{a,2}$, and $g_{a,b}$ show that the coupling capacitance falls off geometrically as the coupling progresses from nearest neighbor to next-to-nearest neighbor and then to next-to-next-to-nearest neighbor coupling, respectively. Data transmons a and b are therefore well isolated from each other, capacitively, by the coupler circuit.

For notational simplicity we will use the shorthand in Table~\ref{table: shorthand} in the following appendices.
\begin{table}
\begin{centering}
\begin{tabular}{c|c}
    term & shorthand\\
\hline
    $g_{C1,2}$ & $g_C$\\
    $g_{Ca,1}$ & $g_a$\\
    $g_{C2,b}$ & $g_b$\\
\end{tabular}
\caption{\label{table: shorthand} Commonly used shorthand notation in the appendices}
\end{centering}
\end{table}

\subsection{Simulation of parametric driving}

We simulated the parametric driving that appears in Fig.~\ref{fig:parametric_coupling} with Eq.~(\ref{eq:HOfullfull}). Every term in Eq.~(\ref{eq:HOfullfull}) is flux sensitive. Those parameters, defined in Table~\ref{table: parametersfull}, are flux sensitive through the definitions of $E_{12}'$ and $E_{j}'$ [Eqs.~(\ref{eq:E12}-\ref{eq:E2})] and $\Phi_{e12}$ [Eq.~(\ref{eq:trueflux})]. We apply a $50\text{ ns}$ half cosine ramp to a flux drive amplitude $A\Phi_0$. The drive $\Phi_e(t) = (A\sin{(\omega_d t)}-0.25)\Phi_0$ was then held constant for $160\text{ ns}$. The qubit state's time evolution was modeled using the qutip python package's time-dependent mesolve ODE solver in the absence of dissipation~\cite{qutip1, qutip2}. Realistic estimates for energy relaxation and decoherence for the proposed tunable coupler are provided in Section~\ref{app:noise}.

\section{Analytical derivation of the effective coupling}
\label{app:SWT}
%
To apply the standard machinery of the Schrieffer-Wolff tranformation to decouple the states of the coupler from the computational Hilbert space we must first diagonalize the coupler states. Exact analytic diagonalization at second order in raising and lowering operators (this neglects anharmonicity which first appears at fourth order in the raising and lowering operators) may be performed using the Bogoliubov transformation \cc{$\Psi = T\Psi'$ where $\Psi^{\dagger} = \begin{pmatrix}a_1 & a_1^{\dagger} & a_2 & a_2^{\dagger}\end{pmatrix}$ is the untransformed ``vector'' of raising and lowering operators. Our goal is to analytically obtain `T'. To this end we define $\Psi^{\dagger}K\Psi$ where $K$ contains prefactors associated with the quadratic terms of the coupler Hamiltonian [Eq.~(\ref{eq:HOfull})]}
\begin{align}
 K &= 
\begin{pmatrix}
\omega_1 & 0 & g_{C}-g_L & g_{C}+g_L\\
0 & \omega_1 & g_{C}+g_L & g_{C}-g_L\\
g_{C}-g_L & g_{C}+g_L & \omega_2 & 0\\
g_{C}+g_L & g_{C}-g_L & 0 & \omega_2
\end{pmatrix}.\label{eq:K}
\end{align}
Note that we use the shorthand notation introduced in Table~\ref{table: shorthand}. \cc{The idea is to perform a similarity transformation $T$ to diagonalize $K$ and thereby find the desired Bogoliubov transformation.}

Interestingly, because $[\Psi_i^{\dagger},\Psi_j]=(-1)^i\delta_{i,j}$, a similarity transformation is likely to produce a set of transformed operators that no longer obey the bosonic commutation relations. We can fix this by identifying the transformation $\sigma = (-1)^i\delta_{i,j}$, and applying it to the column vector $[\Psi_i^{\dagger},\sigma\Psi_j]=\sigma\sigma = \delta_{i,j}$, where the commutation relations are now uniform. A similarity transformation $T$ may be constructed
\begin{align}
\sigma T^{\dagger}\sigma \sigma K T = T^{-1}\sigma K T = \sigma K'
\end{align}
where $K'$ is the desired diagonal form of $K$ and $\Psi = T\Psi'$, see Altland and Simons, page 72~\cite{altland_simons_2010}. Using this technique, the diagonal form of the coupler operators\cc{, given by $\Psi'$,} can be found by diagonalizing $\sigma K$ to obtain the transformation $T$.

\cc{The diagonal elements of $K'$ are a redundant set of eigenenergies $\{\omega_{\pm}, -\omega_{\pm}\}$ otherwise obtainable from the Hamiltonian of the system (again this assumes a perfectly linear system). Starting from Eq.~(\ref{eq:K}) these are,}
\begin{align}
\omega_{\pm} &= \sqrt{2\bar{\omega}^2-\omega_1\omega_2-4g_{C12}g_L\pm 2\eta},
\end{align}
with $\eta = \sqrt{\delta^2\bar{\omega}^2+(g_{C}+g_L)^2\omega_1\omega_2-4g_{C}g_L\bar{\omega}^2}$. 

An analytic form for the eigenstates is also required to obtain an effective coupling but there exists no short form expression for these. Inspection of Eq.~(\ref{eq:K}) reveals that the terms proportional to $g_C+g_L$, the counter-rotating terms, and the terms proportional to $g_C-g_L$, the co-rotating terms, interact with each other at second order in $(g_C\pm g_L)/\omega_j$. This weak interaction between co- and counter-rotating terms justifies separating the calculation of $g_{\textrm{eff}}$ into two parts, greatly simplifying the analytic expression of the eigenstates. The contributions to $g_{\textrm{eff}}$ are then added together to obtain a better approximation of the effective coupling.
%
\subsection{Co-rotating terms}
%
Neglecting terms proportional to $g_{C}+g_L$ in Eq.~(\ref{eq:K}) allows a compact representation of the approximate eigenvectors $T$, which we apply to derive an analytic form of the effective coupling as mediated by exchange interactions. The approximate eigenenergies are likewise simplified to
\begin{align}
\omega_{\pm} = \bar{\omega}\pm\sqrt{\delta^2+(g_C-g_L)^2},
\end{align}
where $\bar{\omega} = (\omega_1+\omega_2)/2$ and $\delta = (\omega_1-\omega_2)/2$. The corresponding eigenvectors now take the form
\begin{subequations}
\begin{align}
a_- &= -\sqrt{1-\beta^2} a_1 +\beta a_2,\\
a_+ &= \beta a_1 + \sqrt{1-\beta^2} a_2,
\end{align}
\end{subequations}
where ${\beta = A/\sqrt{(g_C-g_L)^2+A^2}}$, with ${A = \delta+\sqrt{\delta^2+(g_C-g_L)^2}}$. If $\delta = 0$ then the coupler transmons are degenerate and therefore fully hybridized, leading to $\beta=1/\sqrt{2}$ as we might expect. Substituting the eigenoperators $a_{\pm}$ for $a_{1,2}$ in Eq.~(\ref{eq:HOfull}) and retaining terms to second order, we obtain


%
\begin{align}
H &= H_0 + V
\end{align}
with
\begin{subequations}
\begin{align}
H_0/\hbar &= \omega_a a_a^{\dagger} a_a + \omega_b a_b^{\dagger} a_b + \omega_- a_-^{\dagger} a_- + \omega_+ a_+^{\dagger} a_+,\\
V/\hbar &= 
(-\sqrt{1-\beta^2}g_a +\beta g_{Ca,2})(a_a+a_a^{\dagger})(a_-+a_-^{\dagger})\nonumber\\ 
& \quad +(\beta g_b-\sqrt{1-\beta^2}g_{Cb,1}) (a_b+a_b^{\dagger})(a_-+a_-^{\dagger})\nonumber\\ 
& \quad + (\beta g_a +\sqrt{1-\beta^2}g_{Ca,2})(a_a+a_a^{\dagger})(a_++a_+^{\dagger})\nonumber\\
& \quad +(\sqrt{1-\beta^2}g_b+\beta g_{Cb,1}) (a_b+a_b^{\dagger})(a_++a_+^{\dagger}).
\end{align}
\end{subequations}
We identify a Schrieffer-Wolff generator $S$ such that $[H_0,S] = V$,
\begin{align}
S/\hbar =& \frac{-\sqrt{1-\beta^2}g_a+\beta g_{Ca2}}{\Delta_{a-}}(a_a^{\dagger} a_--a_aa_-^{\dagger})\nonumber\\
+&\frac{\beta g_b-\sqrt{1-\beta^2}g_{Cb,1}}{\Delta_{b-}}(a_b^{\dagger} a_--a_ba_-^{\dagger})\nonumber\\
+&\frac{\beta g_a+\sqrt{1-\beta^2}g_{Ca,2}}{\Delta_{a+}}(a_a^{\dagger} a_+-a_aa_+^{\dagger})\nonumber\\
+&\frac{\sqrt{1-\beta^2}g_b+\beta g_{Cb,1}}{\Delta_{b+}}(a_b^{\dagger} a_+-a_ba_+^{\dagger})\nonumber\\
+& \frac{-\sqrt{1-\beta^2}g_a+\beta g_{Ca2}}{\Sigma_{a-}}(a_a^{\dagger} a_-^{\dagger}-a_aa_-)\nonumber\\
+&\frac{\beta g_b-\sqrt{1-\beta^2}g_{Cb,1}}{\Sigma_{b-}}(a_b^{\dagger} a_-^{\dagger}-a_ba_-)\nonumber\\
+&\frac{\beta g_a+\sqrt{1-\beta^2}g_{Ca,2}}{\Sigma_{a+}}(a_a^{\dagger} a_+^{\dagger}-a_aa_+)\nonumber\\
+&\frac{\sqrt{1-\beta^2}g_b+\beta g_{Cb,1}}{\Sigma_{b+}}(a_b^{\dagger} a_+^{\dagger}-a_ba_+),\nonumber\\
\end{align}
where $\Delta_{j\pm} = \omega_j - \omega_{\pm} = \Delta_j \pm \sqrt{\delta^2+(g_C-g_L)^2}$, $\Sigma_{j\pm} = \omega_j + \omega_{\pm} = \Sigma_j \pm \sqrt{\delta^2+(g_C-g_L)^2}$, $\Sigma_j = \omega_j+\bar\omega$, and as defined in the main text $\Delta_j = \omega_j-\bar\omega$. Then 
\begin{align}
H' = e^{S}He^{-S} = H_0 + [S,V]/2 + \mathcal{O}(V^3)
\end{align}
is the leading order diagonalized Hamiltonian with the pre-factor of the lowest order term $[S,V]/2$ setting the effective coupling
\begin{align}
g_{\textrm{eff}}^{\textrm{co}} =& \frac{(g_ag_b+g_{Ca,2}g_{Cb,1})\beta\sqrt{1-\beta^2}}{2}\nonumber\\
\times&\sum_{j=a,b}\Big[\frac{1}{\Delta_{j+}}-\frac{1}{\Delta_{j-}}+\frac{1}{\Sigma_{j+}}-\frac{1}{\Sigma_{j-}}\Big]\nonumber\\
+&\frac{g_{Ca,2}g_b}{2}\nonumber\\
\times&\sum_{j=a,b}\Big[\frac{1-\beta^2}{\Delta_{j+}}+\frac{\beta^2}{\Delta_{j-}}+\frac{1-\beta^2}{\Sigma_{j+}}+\frac{\beta^2}{\Sigma_{j-}}\Big]\nonumber\\
+&\frac{g_{Cb,1}g_a}{2}\nonumber\\
\times&\sum_{j=a,b}\Big[\frac{\beta^2}{\Delta_{j+}}+\frac{1-\beta^2}{\Delta_{j-}}+\frac{\beta^2}{\Sigma_{j+}}+\frac{1-\beta^2}{\Sigma_{j-}}\Big]\nonumber\\
\end{align}
We may further condense terms
\begin{align}
    g_{\textrm{eff}}^{\textrm{co}} =&(g_a g_b+g_{Ca,2}g_{Cb,1}) (g_C-g_L)\sum_{j=a,b}\Big[\frac{1}{2D_j^2}+\frac{1}{2S_j^2}\Big]\nonumber\\
    +&g_{Ca,2}g_b\sum_{j=a,b}\Big[\frac{\omega_j-\omega_2}{2D_j^2}+\frac{\omega_j-\omega_2}{2S_j^2}\Big]\nonumber\\
    +&g_{Cb,1}g_a\sum_{j=a,b}\Big[\frac{\omega_j-\omega_1}{2D_j^2}+\frac{\omega_j-\omega_1}{2S_j^2}\Big]\\
    D_j^2 &= \Delta_j^2 - \delta^2 - (g_C-g_L)^2\\
    S_j^2 &= \Sigma_j^2 - \delta^2 - (g_C-g_L)^2.
\end{align}
The terms proportional to $S_j^{-2}$ and $g_{Ca,2}g_{Cb,1}$ are typically small and may be neglected.

The ac Stark shifts on the data qubits may be calculated similarly. Here we neglect the terms proportional to $g_{Ca,2}$ and $g_{Cb,1}$
\begin{align}
    \omega^{\textrm{ac}}_a = \frac{g_a^2}{2}\Big[&\frac{\beta^2}{\Delta_{a+}}+\frac{1-\beta^2}{\Delta_{a-}}-\frac{\beta^2}{\Sigma_{a+}}-\frac{1-\beta^2}{\Sigma_{a-}}\Big]\nonumber\\
    \omega^{\textrm{ac}}_a = \frac{g_a^2}{2}\Big[&\frac{\omega_a-\omega_1}{D_a^2}+\frac{-\omega_a-\omega_1}{S_a^2}\Big]\nonumber\\
    \omega^{\textrm{ac}}_b = \frac{g_b^2}{2}\Big[&\frac{1-\beta^2}{\Delta_{b+}}+\frac{\beta^2}{\Delta_{b-}}-\frac{1-\beta^2}{\Sigma_{b+}}-\frac{\beta^2}{\Sigma_{b-}}\Big]\nonumber\\
    \omega^{\textrm{ac}}_b =
    \frac{g_b^2}{2}\Big[&\frac{\omega_b-\omega_2}{D_b^2}+\frac{-\omega_b-\omega_2}{S_b^2}\Big].\nonumber\\
\end{align}
%
\subsection{Counter-rotating terms}
%
Neglecting now the terms proportional to $g_C-g_L$ in Eq.~(\ref{eq:K}), the direct exchange interaction terms, we can calculate the approximate eigenenergies, 
\begin{align}
\omega_{\pm} = \pm\delta+\sqrt{\bar{\omega}^2-(g_C+g_L)^2},
\end{align}
and the eigenvectors,
\begin{subequations}
\begin{align}
a_+ &= \sqrt{1 + \alpha^2} a_1 - \alpha a_2^{\dagger},\label{eq:ap}\\
a_- &= \sqrt{1 + \alpha^2} a_2 - \alpha a_1^{\dagger},\label{eq:am}
\end{align}
\end{subequations}
with ${\alpha = (g_C+g_L)/\sqrt{-(g_C+g_L)^2+D^2}}$, ${D = \bar{\omega}+\sqrt{\bar{\omega}^2-(g_C+g_L)^2}}$. The coefficients of Eqs.~(\ref{eq:ap}-\ref{eq:am}) satisfy the bosonic commutation relations $[a_{\pm},a^{\dagger}_{\pm}] = 1$ and $[a_{+},a_{-}] = 0$.
Substituting the transformed operators for $a_{1,2}$ in Eq.~(\ref{eq:HOfull}) leads to the interaction,
\begin{align}
V/\hbar &=
\alpha g_a (a_a+a_a^{\dagger})(a_-+a_-^{\dagger})\nonumber\\
&\quad+\sqrt{1+\alpha^2} g_b (a_b+a_b^{\dagger})(a_-+a_-^{\dagger})\nonumber\\ 
& \quad+\sqrt{1+\alpha^2} g_a (a_a+a_a^{\dagger})(a_++a_+^{\dagger})\nonumber\\
& \quad+\alpha g_b (a_b+a_b^{\dagger})(a_+-a_+^{\dagger}),
\end{align}
%
%
which can be transformed using the generator,
\begin{align}
S/\hbar &=
\frac{\alpha g_a}{\Delta_{a-}}(a_a^{\dagger} a_--a_aa_-^{\dagger})+\frac{\sqrt{1+\alpha^2} g_b}{\Delta_{b-}}(a_b^{\dagger} a_--a_ba_-^{\dagger})\nonumber\\
+&\frac{\sqrt{1+\alpha^2} g_a}{\Delta_{a+}}(a_a^{\dagger} a_+-a_aa_+^{\dagger})+\frac{\alpha g_b}{\Delta_{b+}}(a_b^{\dagger} a_+-a_ba_+^{\dagger})\nonumber\\
+& \frac{\alpha g_a}{\Sigma_{a-}}(a_a^{\dagger} a_-^{\dagger}-a_aa_-)+\frac{\sqrt{1+\alpha^2} g_b}{\Sigma_{b-}}(a_b^{\dagger} a_-^{\dagger}-a_ba_-)\nonumber\\
+&\frac{\sqrt{1+\alpha^2} g_a}{\Sigma_{a+}}(a_a^{\dagger} a_+^{\dagger}-a_aa_+)+\frac{\alpha g_b}{\Sigma_{b+}}(a_b^{\dagger} a_+^{\dagger}-a_ba_+),\nonumber\\
\end{align}
to obtain the effective qubit-qubit coupling,
%
\begin{align}    
    g_{\textrm{eff}}^{\textrm{counter}} = \frac{\alpha\sqrt{1+\alpha^2}g_ag_b}{2}\sum_{j=a,b}\left[\frac{1}{\Delta_{j-}}+\frac{1}{\Delta_{j+}}-\frac{1}{\Sigma_{j-}}-\frac{1}{\Sigma_{j+}}\right]\label{eq:gcounter}
\end{align}
Approximating $\alpha\sqrt{1+\alpha^2}\approx (g_C+g_L)/2\bar{\omega}$ we may further condense the terms in Eq.~(\ref{eq:gcounter})
\begin{align}    
    g_{\textrm{eff}}^{\textrm{counter}} = -g_ag_b(g_C+g_L)\sum_{j=a,b}\Big[&\frac{\Delta_j/\bar{\omega}}{2D_j^2}-\frac{\Sigma_j/\bar{\omega}}{2S_j^2}\Big].
\end{align}
%

%
\subsection{Effective coupling used in the main text}

We approximate the effective coupling for four coupled harmonic oscillators (remember that all the previous analysis in this section neglects non-linearity) as the sum of the coupling contributions from the co- and counter-rotating terms $g_{\textrm{eff}}^{\textrm{HO}}=g_{\textrm{eff}}^{\textrm{co}}+g_{\textrm{eff}}^{\textrm{counter}}$,
\begin{align}
    g_{\textrm{eff}}^{\textrm{HO}} =\sum_{j=a,b} \Big[&\frac{g_a g_b \left[g_C-g_L-(g_C+g_L)\frac{\Delta_j}{\bar{\omega}}\right]}{2D_j}\nonumber\\
    +&g_{Ca,2}g_b\sum_{j=a,b}\frac{\omega_j-\omega_2}{2D_j^2}\nonumber\\
    +&g_{Cb,1}g_a\sum_{j=a,b}\frac{\omega_j-\omega_1}{2D_j^2}\Big]
\end{align}
We found $g_{\textrm{eff}}^{\textrm{HO}}$ approximates the numerically determined coupling to several percent accuracy when $\nu_j^{(4)},\,\nu_j^{(6)},\,\mu_k^{(4)},\,\mu_k^{(6)} = 0$

Our estimate of $g_{\textrm{eff}}$ when $\nu_j^{(4)},\,\nu_j^{(6)},\,\mu_k^{(4)},\,\mu_k^{(6)} \ne 0$ can be improved by incorporating the contribution of terms forth order in the raising and lowering operators of H in Eq.~(\ref{eq:HOfullfull}) at second order in the raising and lowering operators. We then re-scale terms previously defined in $g_{\textrm{eff}}^{\textrm{HO}}$.

\begin{subequations}
\begin{align}
    &\omega_{a,b}' = \omega_{a,b} - 12\nu_{a,b}\\
    &\omega_{1,2}' = \omega_{1,2} - 2\mu_2^{(4)} - 12\nu_{1,2}\\
    &\bar{\omega}' = (\omega_1'+\omega_2')/2\\
    &\bar{\delta}' = (\omega_1'-\omega_2')/2\\
    &\Delta_j' = \omega_j' - \bar{\omega}'\\
    &g_L' = g_L -3(\mu_1^{(4)}+\mu_3^{(4)})\\
    &D_j^{2'} = \Delta_j^{2'}+\delta^{2'}+(g_C-g_L')^2
\end{align}
\end{subequations}
With these substitutions, we obtain the definition of the effective coupling used in Fig.~\ref{fig:effectivecoupling} of the main text.
\begin{align}
    g_{\textrm{eff}} =&\sum_{j=a,b} \frac{g_a g_b \left[g_C-g_L'-(g_C+g_L')\frac{\Delta_j'}{\bar{\omega}'}\right]}{2D_j'}\nonumber\\
    &\quad+g_{Ca,2}g_b\sum_{j=a,b}\frac{\omega_j'-\omega_2'}{2D_j^{2'}}\nonumber\\
    &\quad+g_{Cb,1}g_a\sum_{j=a,b}\frac{\omega_j'-\omega_1'}{2D_j^{2'}}\label{app:effg w anh}
\end{align}
%


\subsection{Effective parametric coupling used in the main text}

For the purposes of calculating $g_{\textrm{eff}}^p$, we evaluate all the parameters of Table~\ref{table: parametersfull} at $\Phi_{e12} = -0.25\Phi_0$ \textit{except} $g_C$ and $E_{12}'=E_{12}\cos{(\Phi_{e12}/\varphi_0)}$ in $g_L$. We substitute $\Phi_{e12} = A \sin{(\omega_d t)}\Phi_0-0.25\Phi_0$ into the latter term. $\Phi_{e} \approx \Phi_{e12}$ is an approximation of Eq.~(\ref{eq:trueflux}) that is valid in the regime $E_{12}\ll E_j$. The drive flux offset $-0.25\Phi_0$ effectively transforms the cosine into a sine, which we expand about small `A'
\begin{align}
    &\sin{(A \Phi_0\sin{(\omega_d t)}/\varphi_0)}\nonumber\\
    &\quad\quad \approx 2\pi A\sin{(\omega_d t)}-(2\pi A)^3\sin^3{(\omega_d t)}/6\\
    &\quad\quad \approx (2\pi A - (2\pi A)^3/8)\sin{(\omega_d t)}\\
    &\sin^3{(\omega_d t)} = \frac{3}{4}\sin{(\omega_d t)} - \frac{1}{4}\sin{(3\omega_d t)}.
\end{align}
Substituting into $g_L$ we obtain for $g_L^p$
\begin{align}
    g_L &\approx E_{12}(2\pi A - (2\pi A)^3/8)\sin{(\omega_d t)}\sqrt{Z_1Z_2}/2\varphi_0^2\\
    g_L^p &\approx (2\pi A - (2\pi A)^3/8)g_L(0\Phi_0)/2.
\end{align}

The terms proportional to $\sin{(\omega_d t)}$ are large when they include $g_L$ and $\mu_k^{(4)}$ in the numerator of Eq.~(\ref{app:effg w anh}). By contrast, $g_C$ changes very little as a function of flux and may be set to zero. Most other parameters in Table~\ref{table: parametersfull} can be approximately modeled as their static value at $\Phi_{e12} = -0.25\Phi_0$. In the denominator we retained terms without time dependence. This motivates the substitution $(g_C-g_L)^2\rightarrow (g_L^p)^2/2 + g_C^2$.

\begin{align}
    g^p_{\textrm{eff}} \approx \sum_{j=a,b} \frac{-g_a g_b g_L^p(1-\Delta^p_j/\bar{\omega}^p)}{4(D_j^p)^2}.\label{app:parametric}\\
    (D_j^p)^{2} = (\Delta_j^{p})^2 - (\delta^p)^{2} - (g_L^p)^2/2 - g_C^2
\end{align}


\section{Noise analysis}
\label{app:noise}
%
\subsection{Coherence}
The Gaussian pure dephasing rate due to $1/f$ flux noise is $\Gamma_{\phi}^{E} = \sqrt{A_{\Phi}\text{ln} 2}|\partial\omega/\partial\Phi|$ for Hahn echo measurements. It depends upon the noise amplitude $\sqrt{A_{\Phi}}$, defined at $\omega/2\pi = 1\text{ Hz}$, of the power spectral density $S(\omega)_{\Phi}=A_{\Phi}/|\omega|$ and the slope of the energy dispersion with flux $\hbar\partial \omega/\partial \Phi$. Careful engineering gives a noise amplitude $\sqrt{A_{\Phi}} \sim 2.5\mu\Phi_0$.

On the coupler, the peak to peak difference in the frequency dispersion is $\sim 0.6 \text{ GHz}$, giving $\Gamma_{\phi}^{E} \approx 2.5\mu \Phi_0 \sqrt{\text{ln}(2)}(2\pi)^2 \times 0.3 \text{ GHz}/\Phi_0 = 1/40\text{ }\mu \text{s}$. The noise on $g_{\textrm{eff}}$ from the flux sensitivity on $g_L$ is then reduced by a factor $g_ag_b/(2\Delta_j^2-2\delta^2)$. Similarly, frequency noise on the qubits from the flux sensitivity on $\bar \omega$ is reduced by a factor of $\sim g_a^2/(2\Delta_j^2-2\delta^2)$. For $g_{\textrm{eff}}=2\pi\times 60\text{ MHz}$, the coupler limits the pure dephasing lifetime to $T_{\phi}^E \sim 400\text{ }\mu\text{s}$.
%
\subsection{Energy relaxation}
%
Energy relaxation of a qubit into its nearest neighbor coupler transmon is approximately given by the Purcell formula. Using the definitions in the main text and supplement the coupler induces relaxation of qubit a
\begin{align}
    \Gamma_{1,a}^{\textrm{Purcell}} &\approx g_a^2\left(\frac{\Gamma_{1,+}\beta^2}{\Delta_{a+}^2}+\frac{\Gamma_{1,-}(1-\beta^2)}{\Delta_{a-}^2}\right)\nonumber\\
    &\sim \frac{\Gamma_{1,1} g_a^2}{(\omega_a-\omega_1)^2}.
\end{align}
Similarly, for qubit b
\begin{align}
    \Gamma_{1,b}^{\textrm{Purcell}} &\approx g_b^2\left(\frac{\Gamma_{1,-}\beta^2}{\Delta_{b-}^2}+\frac{\Gamma_{1,+}(1-\beta^2)}{\Delta_{b+}^2}\right)\nonumber\\
    &\sim \frac{\Gamma_{1,2} g_b^2}{(\omega_b-\omega_2)^2}.
\end{align}
The energy relaxation rates for coupler transmons 1 and 2 are given by $\Gamma_{1,1}$ and $\Gamma_{1,2}$, respectively. While these quantities are not true observables of the system since the coupler transmons can be strongly hybridized, we want to emphasize that qubit a is not very sensitive to relaxation channels local to coupler transmon 2, nor is qubit b sensitive to relaxation channels local to coupler transmon 1. In the case of direct measurement of coupler relaxation rates, or if we expect correlated relaxation processes, then $\Gamma_{1,+}$ and $\Gamma_{1,-}$ describe the energy relaxation rate of the upper and lower hybridized states, respectively.
\par
Plugging in $20\text{ }\mu\text{s}$ as a reasonable lower bound on each coupler transmon's energy relaxation lifetime, the induced $T_{1}$ on a neighboring data qubit is $220\text{ }\mu\text{s}$ for a $g/\Delta$ ratio of $0.3$.
%
\subsection{Coupling a qubit to an open quantum system}
%
We consider coupling qubit `$a$' to a bath as mediated by the tunable coupler. In this scenario a deliberate interaction with the bath induces coupler transmon 2 to relax with rate $\Gamma_{1,2}$ into the bath.
\begin{align}
    \Gamma_{1,a} \approx& \Gamma_{1,2} g_a^2\beta^2(1-\beta^2)\left(\frac{1}{\Delta_{a+}^2}+\frac{1}{\Delta_{a-}^2}\right)\nonumber\\
    \approx& \frac{\Gamma_{1,2} (g_C-g_L)^2g_a^2}{(\Delta_a^2-\delta^2-(g_C-g_L)^2)^2}
\end{align}
We see that the coupler isolates the qubit from dissipation on the next-to-nearest-neighbor coupler transmon to fourth order in $g_C-g_L,\,g_a/\Delta_a$. Although $\Gamma_{1,a}$ turns off at $g_C-g_L = 0$, it is difficult to achieve sizeable `on' $\Gamma_a$ values using the coupler in dispersive operation. This weak `on' interaction motivates the alternative approach taken in the main text.
%
\subsection{Energy relaxation into the flux bias line}
%
A critical consideration for choosing an appropriate mutual inductance between the coupler SQuID and its flux bias line is  the relaxation rate of the coupler induced due to this inductive coupling. This relaxation can lead to strong correlations if $\delta \ll |g_C-g_L|$. In this circumstance the `bright' state is the one that tunes strongly with flux and will relax, at worst, at the sum of the individual transmon relaxation rates. 
In the other regime $\delta \gg |g_C-g_L|$, the eigenstates $\ket{\pm}$ are closely approximated by independent transmon eigenstates, such that we can approximately map $\xi_{j}\leftrightarrow \xi_{\pm}$ for $j\in \{1,\,2\}$.
Assuming $\dot{\Phi}_e = M \dot{I}$, this allows us to write the effective decay rate as,
\begin{align}
\Gamma_{1,\pm}^{fb} 
&\sim \left(\frac{\xi_{\pm}(\Phi_e)M}{\Phi_0}\right)^2 S_{\dot{I}\dot{I}}(\omega)\nonumber\\
&= \frac{2\hbar\xi_{\pm}^2(\Phi_e)M^2\omega_{\pm}^3}{\Phi_0^2 Z_0},
\label{eq:test}
\end{align}
where in the first line we have used the flux participation ratio as prescribed by the third line of Eq.~(\ref{eq:HOfull}), and in the second line we have assumed that the magnitude of current fluctuations is set by their vacuum expectation value, i.e. $S_{\dot{I}\dot{I}}(\omega) = \omega_{\pm}^2S_{II}(\omega) = 2\hbar\omega_{\pm}^3/Z_0$.
\par
Given a mutual inductance $M = 3.7\textrm{ pH}$, capacitive coupling $C_{12}\ll C_1,\,C_2$, dimensionless coupling constant 
\begin{align}
    \xi_{\pm}(0) &\sim (\Phi_0/\sqrt{2\hbar Z_1})E_{12}'(0)/E_1'(0)\nonumber\\
    &\textrm{or }(\Phi_0/\sqrt{2\hbar Z_2})E_{12}'(0)/E_2'(0),\nonumber
\end{align}
transition frequency $\omega_{\pm}/2\pi = 5\textrm{ GHz}$, and bath impedance of $50\textrm{ Ohms}$, the equation above leads to an estimated $\Gamma_{1,\pm}^{fb}\sim 1\times 10^3\textrm{ s}^{-1}$, before additional low pass filtering of the flux bias. We note that these are worst case calculations since $\xi_{\pm}(\Phi_e)\propto \cos{(2\pi\Phi_e/\Phi_0)}$, which at $\Phi_e^{(0)} \sim 0.25 \Phi_0$ causes the dimensionless coupling constant $\xi(\Phi_e)$ to vanish.
\bibliographystyle{apsrev4-2}
\bibliography{Main}
\end{document}